\documentclass[conference,compsoc]{IEEEtran}
\usepackage{makecell}
\usepackage{multirow}
\usepackage{rotating}
\usepackage{tablefootnote}
\usepackage[table]{xcolor}
\usepackage{comment}
\usepackage{url}
\usepackage{graphicx}
\usepackage{caption}
\usepackage{standalone}
\usepackage{tikz}
\usepackage{amsmath}
\usepackage{todonotes}
\usepackage[most]{tcolorbox}
\usepackage{booktabs}
\usetikzlibrary{arrows.meta, positioning}
\usepackage{longtable}
\newtcolorbox{takeaways}{
  colback=gray!15,
  colframe=gray!50,
  arc=4pt,
  boxrule=0pt,
  left=6pt,
  right=6pt,
  top=6pt,
  bottom=6pt,
  before skip=0.8\baselineskip,
  after skip=0.8\baselineskip
}

\usepackage{chngcntr}

\ifCLASSOPTIONcompsoc
  \usepackage[nocompress]{cite}
\else
  \usepackage{cite}
\fi

\ifCLASSINFOpdf
\else
\fi

\begin{document}
\title{A Meta-Study on Replication Papers in Usable Security \& Privacy}

%\author{\IEEEauthorblockN{Name}
%\IEEEauthorblockA{Info}
%\and
%\IEEEauthorblockN{Name}
%\IEEEauthorblockA{Info}
%\and
%\IEEEauthorblockN{Name}
%\IEEEauthorblockA{Info}}

% compsoc conferences compared to traditional conferences), use this
\author{\IEEEauthorblockN{Christian Mack\IEEEauthorrefmark{1},
Benjamin Maximilian Berens\IEEEauthorrefmark{1},
Hanna Algedri\IEEEauthorrefmark{1}, 
Tobias Hilt\IEEEauthorrefmark{1}, \\
Daniela Reimer\IEEEauthorrefmark{1},
Peter Mayer\IEEEauthorrefmark{1}\IEEEauthorrefmark{2} and
Melanie Volkamer\IEEEauthorrefmark{1}}
\IEEEauthorblockA{\IEEEauthorrefmark{1}Karlsruhe Institute of Technology}
\IEEEauthorblockA{\IEEEauthorrefmark{2}University of Southern Denmark}}

\maketitle

\begin{abstract}
The field of usable security and privacy research is a young and expanding field, which is still developing standards for its research, e.g. regarding replications. %One underdeveloped standard is the replication of prior work showcased by the limited amount of replication and wide variety of interpretation of the term replication. 
We used a mixed-method approach, in order to get a better understanding of the current state of replications in the field of usable security and privacy: (1) we examine the Call for Papers of 13 venues spanning security, privacy, and human-computer interaction; (2) we conduct a systematic search for papers reporting replicated user studies published across these venues between 2016 and 2025, yielding 24 relevant publications; (3) 
we categorized these 24 papers employing the replication taxonomy proposed by Olszewski et al. (2025); (4) we distributed a survey to the authors of these papers to understand their motivations for conducting replications.
Our analysis reveals four key insights: (A) Calls for Papers would benefit from clearer guidelines for authors and reviewers regarding replication work; (B) determining what modifications were made relative to the original study proves difficult when reading replication papers; (C) strict exact replications do not exist in our sample. Approximately two-thirds of the 24 studies altered multiple aspects of the original work; (D) temporal and contextual changes affecting results emerged as one of the most frequently cited motivations for replication. % \textcolor{green}{Warum das eigentlich, wenn andere Gruppe mehr Nennung? Weil das innerhalb der Gruppe homogener war? Finde nur das fällt auf, wenn man sich Table 4 anschaut.}. 
Based on these findings, we offer practical recommendations for venues, researchers, and peer reviewers to strengthen replication practices in usable security and privacy research.

\end{abstract}


% \fi
\IEEEpeerreviewmaketitle

\section{Introduction}  \label{intro}

Replicating research results is widely recognized as a cornerstone of the scientific process. The general idea of replication research is to allow researchers to independently verify each other's findings, complementing peer-reviews as a safeguard that findings and conclusions are robust and trustworthy. The usable security and privacy community widely introduced the category of replication papers as a dedicated type of paper through the 2017 Call for Papers (CfPs) of their main venue, SOUPS (Symposium of Usable Privacy and Security)\footnote{\url{https://www.usenix.org/conference/soups2017/call-for-papers}}. Workshops such as EuroUSEC (European Workshop on Usable Security) and USEC (Workshop on Usable Security) followed suit and started encouraging submissions of replication studies in their Call for Papers\footnote{See, e.g.\url{https://www.ieee-security.org/TC/EuroSP2016/cfp.php}}.
Given the fact that the usable security and privacy community acknowledged the importance of replication research already 10 years ago, we started this work with the assumption that this acknowledgment of importance should have been translated into publications replicating previous work.

Therefore, in this work, we investigate the current prevalence and practice of replications in the area of usable security and privacy. We use this systematization to propose guidance for the usable security and privacy community on what to replicate, what might be best practices, how to describe replications, as well as how to review replication papers. 
We employ a framework to classify replication studies for the following reasons: First, it reveals patterns in how authors understand and conceptualize replication. Second, it contributes toward a common understanding of replications by identifying typical similarities and differences. Third, it allows us to evaluate the framework's utility and limitations for categorizing replications in future research.

The focus of our investigation is on the following 13 venues: \textit{CHI, SOUPS, USEC, EuroUSEC, CCS, CSCW, ICSE, PETS, S\&P, EuroS\&P, NDSS, WWW, and USENIX Security}. This selection includes the venues in the analysis by Klemmer et al. \cite{klemmer_how_nodate}, who examined papers in the usable security and privacy field in terms of the availability of artifacts and details about their study designs.

We complement this list with USEC and EuroUSEC. 

%, 2) to learn from the usable security and privacy community for   , i.e. what type of replications are conducted.  %; and (2) %To that end we performed a systeamtic literature review 

In our investigation, we conducted a wide variety of analyses in a four-pronged approach. Firstly, we evaluated the Call for Papers (CfPs) of the 13 focus venues to identify if they solicit replications and, if so, in which manner, seeking to answer our first research question:

\newcommand{\rqspacing}{4pt}
\newcommand{\rqcfp}{How is replication research defined according to the Call for Papers across prominent venues in the field of usable security and privacy?}
\vspace{\rqspacing}
\noindent \textbf{RQ1:} \textit{\rqcfp}

\vspace{\rqspacing}
\noindent{}
We noticed that from those 13 venues only about half encourages replications explicitly while only SOUPS provides some specifications for replications, i.e., ``confirm, question, or clarify results while clearly describe the methodological differences and compare their own with original results''. Furthermore, study ``protocols can be the same as the original study or may vary one or more key variables [...]''.

Second, we identified replication papers in the proceedings of the 13 focus venues, thereby answering our second research question:

\vspace{\rqspacing}
\newcommand{\rqprevalence}{How prevalent are replications in the field of usable security and privacy across prominent venues of this field?}
\noindent \textbf{RQ2:} \textit{\rqprevalence}

\vspace{\rqspacing}
\noindent Based on a systematic literature review, we identified replication papers in the proceedings of the 13 venues mentioned above, mapping the prevalence of replications in general and in different venues. Overall, We could only identify 24 replication papers in the time frame from 2016 to 2025, making replication in this field rare. Almost half of those were published at SOUPS which seems to establish this venue as the dominant one for replication work. \\ %This seems to be supported by our analysis of the CfPs of these venues, were SOUPS emerged as the first venue that solicited the submission of replication papers in as a dedicated type of submission.
Third, we classified the $n=24$ replication papers using the replication framework from Olszewski et al.~\cite{olszewski_sok_nodate}, aiming to provide insight for our third research question:

\vspace{\rqspacing}
\newcommand{\rqreplicationtypes}{Of what types are the replication studies in the usable security and privacy field?}
\noindent \textbf{RQ3:} \textit{\rqreplicationtypes}

\vspace{\rqspacing}
\noindent %As theoretical foundation, we used the replication framework by Olszewski et al.~\cite{olszewski_sok_nodate}. 
Since Olszewski et al.'s framework had not been used for empirical user studies before, we adapted it for that context. % slightly (adopting to context was explicitly mentioned by Olzewski). % by merging the method and data layers of their original framework. 
Our adaption takes into account the fact that in empirical user studies, the data collected is directly inter-related to the method used. The adapted framework was used to classify the 24 papers. All were classified as conceptual replications (see Figure \ref{validityTree2}), with two thirds changing more than one aspect of the original paper. Most papers  changed the method and / or the analysis.

The adopted framework %can be seen as 
represents an additional contribution of this paper.

Last but not least, we conducted an online survey among the authors of the previously identified replication papers in order to empirically identify motivators and hindrances for replications, thereby answering our fourth research question:

\vspace{\rqspacing}
\newcommand{\rqsurvey}{What are motivators to conduct replication studies?}
\noindent \textbf{RQ4:} \textit{\rqsurvey}

\vspace{\rqspacing}
\noindent %In our survey of replication paper authors, 
We found that confirming, extending, or generalizing earlier results emerged as the most prominent motivators for conducting replication studies. In particular, confirming whether the results still hold over time as mental models change emerged as an important theme. Many respondents also cited the desire to save resources and time (e.g., harness existing artifacts, study materials, and methods) as a practical incentive. % When asked what types of studies should be replicated the majority of participants indicated in their responses that any study irrespective of its design or domain should be replicated.

\ \\
From the answers to these research questions, 
%From the various investigations, we deduced open questions for the usable security and privacy community. 
we derive \textit{recommendations} for authors, reviewers, and venues. We provide a table, authors of future replications studies could use to explain and justify the differences between their research and the original study they replicate.

\section{Related Work}
%This section reviews related work on replication and transparency, presents our analysis of Calls for Papers, and introduces a framework for classifying replications.

%\subsection{Replication and Transparency}
\textbf{Motivation to replicate. } Several papers motivate the need for replications independent of the concrete research domain (e.g. ~\cite{cacioppo2015social}), and in particular the need for multiple replications to identify generalizable effects rather than context-specific findings~\cite{Maxwell2015}. % and (2) the need to replicate in none-WEIRD (Western, Educated, Industrialized, Rich and Democratic) countries as WEIRD countries are the primary focus of many empirical studies~\cite{hasegawa_how_2024}. 
Thus, we focus on ``replications'' which aim to challenge the (generalizability of) the original work's findings, rather than reusing parts of its methodology to answer different research questions.
\\\\
\textbf{Replication definitions. } Several papers across research disciplines have noticed challenges in defining what constitutes a replication (see e.g.,~\cite{gomez_replications_2010}); % and systematic errors in the original paper's design that limit the replicability altogether~\cite{sotirakopoulos_challenges_2011}.
including inconsistent terminology such as \textit{replicate} and \textit{reproduce}~\cite{goodman_what_2016}. This terminology challenge also holds for the usable security and privacy community; e.g., SOUPS differentiates in the Call for Paper of 2017 between \textit{full replication}, \textit{variation} and \textit{triangulation}, whereas ACM distinguishes \textit{repeatability}, \textit{reproducibility} and \textit{replicability}~\cite{ACM_BadgeArtifacts_2020}.
%%%XXX
Furthermore, while it is widely acknowledged that the degree of similarity between the original study and the replication attempt can differ, there exist different proposals how to categorize the different types of replications. Often it is differentiated (see e.g.{~\cite{makel2012replications,hornbaek2014once,marsden}) between \textit{direct} (or \textit{strict}) replications, which aim to reproduce the original study as closely as possible, \textit{partial} (or \textit{approximate}) replications, which vary  study characteristics while preserving core elements of the original study, and \textit{conceptual} replications, which research the same research questions while using different approaches to study them. 
Recent work argued that these distinctions are overly simplistic and proposed more fine-grained frameworks that categorized replications based on the specific aspects that differ from the original study, {e.g.~\cite{cruz2023model,hoffmann2025design}.
Olszewski et al.  proposed such a framework to distinguish 16 types of replications in computer security~\cite{olszewski_sok_nodate}. 
Note, we summarize their framework in Subsection~\ref{originalframework}. One of our contributions is to adapt and apply this framework for the usable security and privacy context.
\\\\
\textbf{Papers classifying replications.}
Researchers have already  investigated in various fields what type of replications have been conducted: e.g., Makel et al.~\cite{makel2012replications}  reviewed articles published in psychology journals. They  found that most replications were conceptual replications and that there were only very few direct replications. Within Human--Computer Interaction (HCI), Hornb{\ae}k et al.~\cite{hornbaek2014once} categorized replication papers from leading HCI venues. They also found that most of the replications were conceptual.
While their research areas are related to usable security and privacy research, both papers are more than 10 years old and as such were published before researchers proposed   fine-grade categorization frameworks for replication studies. 
\\\\
\textbf{Paper with related goals.} We note that a related meta-analysis on replications in human-centered security has been published at the Workshop MetaCRiSP 2026 during our revision period (Schmüser et al. (2026) \cite{schmuser2026position}). They also investigated the Call for Papers and did a literature search to identify replication papers. However, they did not try to categorize these papers. A detailed comparison of the methodology and results is provided in Section~\ref{PositionPaper}.

\section{RQ1: Replication in Calls for Papers}
\label{CfP}

\begin{figure*}[!ht]
  \centering
  \includegraphics[width=\textwidth]{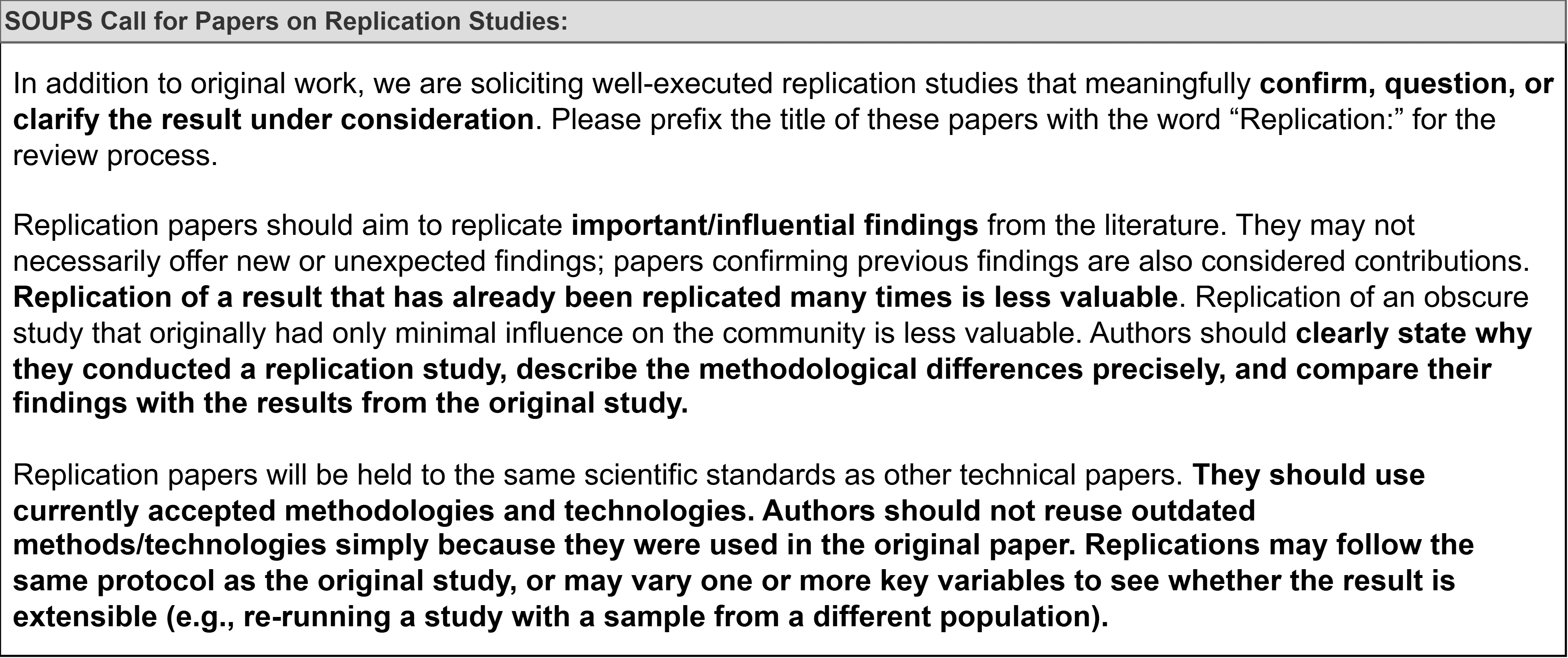} 
  \caption{Call for Paper paragraph about replication studies at the Symposium On Usable Privacy and Security (SOUPS)}
  \label{fig:soupCFP}
\end{figure*}

This section contributes to answer the following research questions.
\textbf{RQ1:} \textit{\rqcfp}
\ \\
% community expresses wish for more transparency and replications as e.g. there were/re dedicated workshops that discuss this --> cfp should represent this wish --> we analyse cfp to see if it matches
 
We wanted to check whether relevant venues ask for replications, and if so, to understand what type of replications they are looking for -- according to their Call for Papers (CfP). 
Thus, we analyzed the Call for Papers (CfP) of the 13 venues\footnote{The explanation for these 13 venues is provided in Section \ref{intro}.}: \textit{CHI, SOUPS, USEC\footnote{We could not find two of the CfP from USEC and EuroUSEC.}, EuroUSEC, CCS, CSCW, ICSE, PETS, S\&P, EuroS\&P, NDSS, WWW, and USENIX Security}. %This list follows Klemmer et al.~\cite{klemmer_how_nodate}, who examined papers in the usable security and privacy field in terms of the availability of artifacts and details on their experiments.~\footnote{~\cite{klemmer_how_nodate} did not include USEC in their list.}
The analyzed CfPs differ in how strongly they encourage replication papers, which we distinguish in the following:

\begin{table*}[!ht] 
  \centering
  \caption{Call for Papers based on their level of encouragement of replication papers.}
  \label{tab:callforpaper}
\begin{tabular}{cc|cc%|c
}
  \toprule
  \multicolumn{2}{c}{\textbf{\makecell{Encouraging Replications\\and Providing some Specifications}}} &
  \multicolumn{2}{c}{\textbf{Encouraging Replications}}  \\ \midrule %  \textbf{\makecell{Transparency Requirements\\as Precondition for future Replications}} \\ \midrule
  Venue & Since & Venue & Since
  %&   %Venues 
  \\ \hline
  ~ & ~ & SOUPS, USEC, EuroUSEC & 2016 
  %& SOUPS, CHI, WWW,   
  \\
  SOUPS & 2017& AsiaUSEC & 2020 %& CCS, CSCW, EuroS\&P,  
 % \\
%  SOUPS & 2017 & CHI & 2021 %& USENIX Security, USEC,  
  \\
  ~ & ~ & USENIX Security & 2024 % &  EuroUSEC, ICSE
  \\
  ~ & ~ & NDSS & 2025 %&  ~
  \\
  %\bottomrule
\end{tabular}
\end{table*}

\textbf{Encouraging Replications and Providing some Specifications: }The SOUPS 2017\footnote{In the SOUPS CfP of 2016, replication studies were mentioned in the list of topics. CfP SOUPS 2016: \url{https://www.usenix.org/conference/soups2016/call-for-papers} CfP SOUPS 2017: \url{https://www.usenix.org/conference/soups2017/call-for-papers}} CfP emphasized replication and distinguished three categories of replications, i.e. the CfP states exactly the following:
\begin{itemize}
    \item  Full replication: Same study protocol, same type of sample.
\item Variation: One design variable is changed. For instance, re-running
an MTurk study with a different sample; conducting a study with a
sample from different countries, etc.
\item Triangulation: Same study goal but different design. For instance,
conducting a field study instead of a self-reporting survey; using a
different measurement instrument to measure a variable
\end{itemize}
The 2018 CfP expanded the description but removed the categorization. We could not find any information on why it was removed. 
Since 2019, the wording has remained stable. The phrasing is provided in Figure \ref{fig:soupCFP} while we highlight in bold the parts relevant for this paper.

 \textbf{Encouraging Replications: } 
SOUPS encouraged replications in 2016. 
 USEC and EuroUSEC  explicitly recognize replication, since at least 2016. AsiaUSEC followed this approach for its singular event in 2020. There is no definition but the CfP encourages replications: e.g. EuroUSEC 2021 says: ``Reports of researchers replicating previously published research studies. We want to encourage the submission of replication of studies, as this serves to mature the science of usable security and privacy research and validates previously published research findings.'' \footnote{\url{https://eurousec.secuso.org/callforpapers.php} Last accessed: August 1st 2026} 

  CHI includes replications as part of empirical research contributions since 2021 in the subcommittee ``Computational Interaction''\footnote{\url{https://chi2021.acm.org/for-authors/presenting/papers/selecting-a-subcommittee} Last accessed: August 1st 2026} but until 2025 it was not included in the subcommittee ``Privacy and Security''. Therefore, for our work, we consider CHI (``Privacy and Security'') as one of the venues which does neither recognize nor even mention replication. 
 
 USENIX Security mentions replication as part of Systematization of Knowledge (SoK) papers in their CfP (``While both SoK and survey papers may involve summarizing existing research, the key difference is that a SoK paper provides a more structured and insightful overview, which might also involve new experiments to replicate and compare previous solutions.''\footnote{\url{https://www.usenix.org/conference/usenixsecurity24/call-for-papers} Last accessed: August 1st 2026})  for the first time in 2024 and continues to include them. NDSS adopted a similar approach in 2025 and also mentions  replications as part of SoK papers. Since 2025 USENIX Security directly encourages replication papers (``We encourage submissions that not only replicate studies but also offer meta-analyses that assess the replicability of research.''\footnote{\url{https://www.usenix.org/conference/usenixsecurity25/call-for-papers}}) %\textcolor{red}{todo gibts hier ne definition}.
 ICSE mentions replications in their CfP in the years 2020 and 2021, but only in passing\footnote{ICSE CfP 2020: \url{https://2020.icse-conferences.org/track/icse-2020-papers?#Call-for-Papers} , ICSE CfP 2021: \url{https://conf.researchr.org/track/icse-2021/icse-2021-papers#Call-for-Papers}}

Interestingly, five of the 13 venues (WWW, CCS, CSCW, EuroS\&P and ICSE) have transparency requirements in their CfP but do not  explicitly mention replication studies as paper type. Table \ref{tab:callforpaper} shows an overview of venues encouraging replications in their CfP.

\section{Pre-Considerations for RQ3: Framework Adaption for User Studies}\label{adoption} 
We first introduce the replication framework from Olszewski et al. (2025)~\cite{olszewski_sok_nodate} as we base our categorization on their framework, then we explain how we instantiated and adapted it for the usable security and privacy context. Lastly, we use a mock example to explain the different type of replications in the adapted framework. 
\subsection{Framework from
 Olszewski et al.} 
 \label{originalframework}

%\textcolor{red}{ist klar, dass die eigentlich was machen wollten um das 'original' paper zu bewerten, welche form von replications möglich sind}
Olszewski et al.~\cite{olszewski_sok_nodate} published their paper titled ``SoK: Towards a Unified Approach to Applied Replicability for Computer Security'' at the USENIX Security Symposium.  Their first contribution is the systematization of research and recommendations related to reproducibility, replicability, and validity. Based on this systematization, they identified several shortcomings of existing %reproducibility and replicability 
frameworks. 
As their second contribution, Olszewski et al. provide a framework for reasoning about replicability called ``Tree of Validity (ToV)''%\footnote{The authors use the term validity to avoid confusion with the various definitions of reproducibility and replicability. }
. Their replication framework allows to classify different forms of replications using a binary tree (see Figure~\ref{fig:replication_framework}), i.e., it allows researchers of replication studies to better contextualize their contribution and their changes. The root of the tree is the most critical aspect, as it assumes the \textbf{problem definition}, i.e. the research question of the original research and its replication, to be identical. %\textcolor{red}{ich finde das Problem wurde hier nicht klar.  Es sollte in dem fall auch "root" und nicht "note" heißen, oder? Ich habs mal umgeschrieben, aber deins hier im Kommentar gelassen}
%The tree’s note is the problem, i.e. the research question which the conducted research is supposed to answer. This means, for research being a replication in terms of their replication framework, it is essential that the problem definition is the same for the original research and its replication. 
In their tree, there are four layers: 

\textbf{Domain:} ``The domain is the environment of the experiment. It can include (but is not limited to) the population studied, time, software systems, hardware systems, etc.''~\cite{olszewski_sok_nodate}

\textbf{Method:} ``A method is the approach to gather and/or manipulate data.''~\cite{olszewski_sok_nodate}
    
\textbf{Data:} ``The data is the collection of measurements or observations within the setting.''~\cite{olszewski_sok_nodate}
   
\textbf{Analysis:} ``Finally, the analysis is conducted on the data and provides a quantifiable measure.''~\cite{olszewski_sok_nodate}

\begin{figure}
    \centering
    \includegraphics[width=0.91\linewidth]{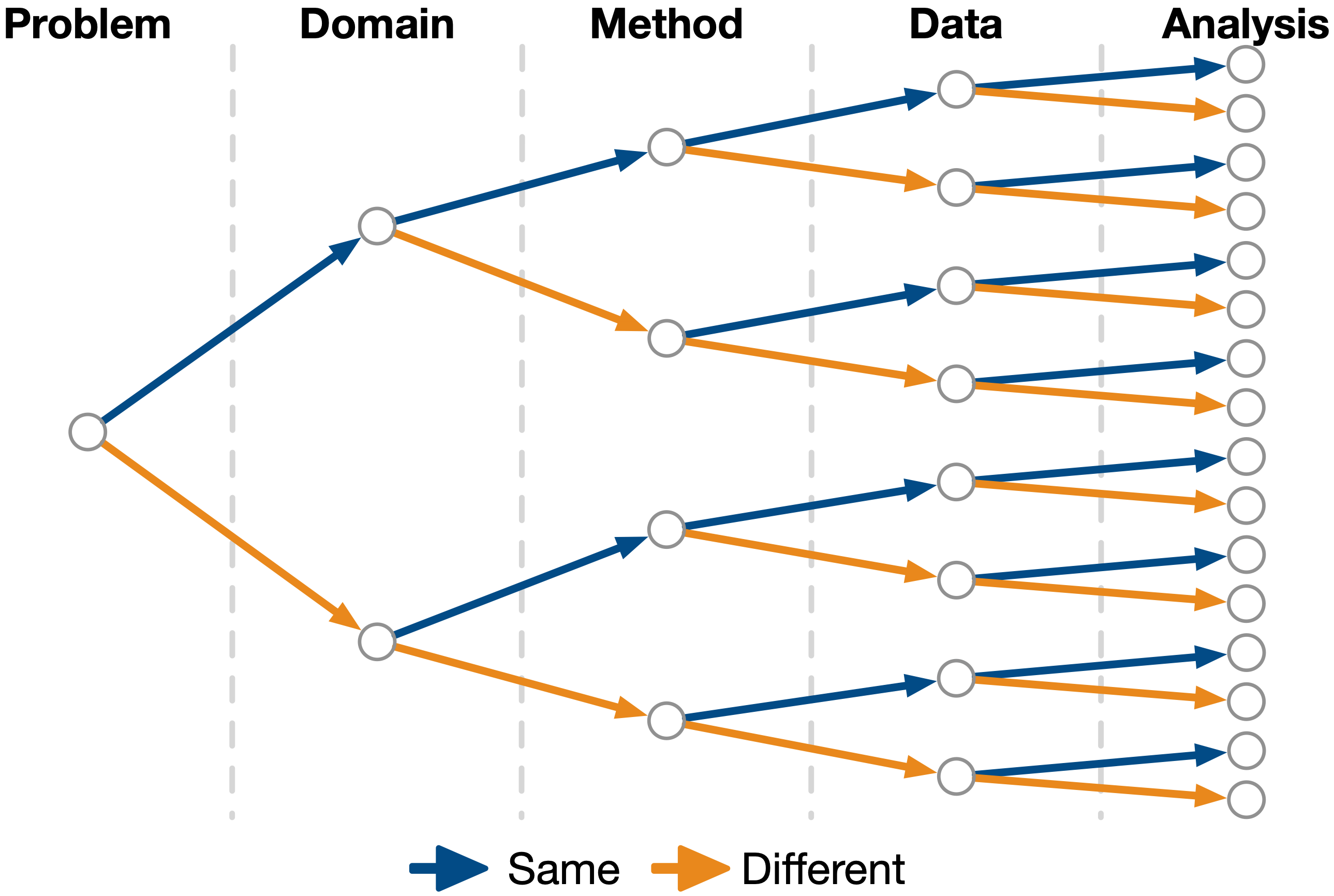}
    \caption{Replication Framework from Olszewski et al.~\cite{olszewski_sok_nodate}}
    \label{fig:replication_framework}
\end{figure}
 Each node of the decision of the binary tree represents a decision point whether the part remains the same or is different for each of the four layers. The authors also discuss how to apply it to different domains. In this regard, they explain that layers can be swapped or removed to make it a better fit for some domains. %For example, in some experiments, it may make sense to move data before method or remove either layer."
Furthermore, Olszewski et al. explain how different replication and reproducibility definitions such as those provided in~\cite{ACM_BadgeArtifacts_2020,nationalacademiesReproducibilityReplicability,gomez_understanding_2014,goodman_what_2016,gunderson_the_2020}  %by ACM~\cite{ACM_BadgeArtifacts_2020}, National Academies of Science~\cite{nationalacademiesReproducibilityReplicability}, Gomez at al.~\cite{gomez_understanding_2014}, Goodman et al.~\cite{goodman_what_2016} or Gunderson at al.~\cite{gunderson_the_2020})  
can be mapped to their framework. The authors explain why their framework of replicability can broadly be applied for, and beyond, security researchers. Although Olszewski et al. do not discuss usable security and privacy research in their paper, they mention that human-computer interaction can use their framework, too. 
%moved hiere from 4.2
The proposed framework aligns with the replication requirements described in the SOUPS CfP. In particular, the requirement to ``meaningfully confirm, question, or clarify the result under consideration'' corresponds to the framework’s problem layer, which is assumed to remain constant across replications. In contrast, the remaining layers may be the same or different, which is consistent with the CfP’s recommendation to employ current methodologies rather than reproducing outdated procedures from the original study. Similarly, the CfP allows (but does not require) adherence to the original protocol and permits modification of one or more key variables. These options correspond to the framework’s binary decisions across its respective layers.
All this motivated us to use their framework as basis for our work.
% and applies a structure for comparing validity studies	

\subsection{Instantiation and Adaptions}
We followed the recommendation by  Olszewski et al.~\cite{olszewski_sok_nodate} and first discussed with all authors whether we want to use the framework as proposed %(see Figure \ref{fig:replication_framework}) 
or adapt it, i.e. either remove a layer and / or change the order of the layer. % We first explain how and why we adapted their framework. We then provide a fictitious user study to illustrate for each type of replication how such a replication could look like. 
Our discussion was based on the replications we could think of, as well as those we conducted ourselves in the past. 
We realized that in the context of user studies, the layer ``method'' and the layer ``data'' are very closely related. Therefore, we combine these layers into one layer which we call ``method'' (see also Figure \ref{validityTree2}). 
Taking every note representing a binary decision into account, there are eight different types of replication in our adopted framework. Note, that for our paper, the first two types (domain and method being the same as the original study) are called \textit{exact replications} and the
remaining six are called \textit{conceptional replications}.

Next, all SOUPS papers with the term replication in the title were read by at least one author. Although, having replication in the title, we noticed that not all would count as replication if we are strict regarding the precondition from Olszewski et al.~\cite{olszewski_sok_nodate}, i.e. 
% We had several discussions regarding the problem definition being the root of the tree. This means, 
the actual problem mentioned in the original paper as well as the one mentioned in the replication paper are the same (e.g. \cite{danilova2020replication,baig2021replication}). %relevant to decide whether a paper is considered as replication or not. 

Therefore, we decided that for being considered as replication, it is sufficient that the same research question can be answered without explicitly naming it.

We also discussed what is part of the different layers. For domain, we identified the following list of examples: different sample, different environment, e.g., on iOS instead of Android and older or new versions of operating systems.
For methods, we identified the following list of examples, e.g., switching from interview to survey, changing phrasing of questions and / or answer options. Different analysis means for instance different statistical tests.%
\subsection{Example} % for Original Study}
\label{example}
 
In this subsection, we describe a simplified artificial original study, which we use to demonstrate how different changes to the study result in classification into the eight replication types. We start by describing the original study.

\textit{Problem:} Does the anti-phishing awareness video ``Phishly'' support people in significantly better distinguishing between phishing and legitimate emails?

    \textit{Domain:} Student sample in U.K. in 2022
    
    \textit{Method:} Online survey with the following main steps:  judging email screenshots (phish yes/no) - participating in Phishly - judging email screenshots 
    
    \textit{Analysis:} Hit, miss, false alarm and correct rejection values from before and after participating in ``Phishly'' for each participant are analyzed using     
    repeated measures ANOVA.\\ % with within subject data from before and after participating in 'Phishly'
\ \\
\textbf{Handling of Problem Definitions.}
%\textcolor{red}{ggf passt das auch in methodik, da wo wir schreiben, das wir grosszügig bei der problembeschreibung waren?}
We discuss some example studies with similar problem definitions to demonstrate the impact of the original problem description:%, i.e. research question: 

{{Problem-1:}} Does the anti-phishing awareness video ``Phishly'' support people in significantly better distinguish between phishing and legitimate emails \textit{even when being under time pressure}?

%\textbf{Problem-2:}  Does the anti-phishing awareness video 'Phishly' supports people in significantly better distinguishing between phishing and legitimate emails \textit{1 year after having watched this video}?

Problem-2: Does the anti-phishing awareness video ``Phishly'' support \textit{elderly people} in significantly better distinguishing between phishing and legitimate emails?

Problem-3:  Does the anti-phishing awareness \textit{game ``Phish-ME''} support people in significantly better distinguishing between phishing and legitimate emails?
    \\
    We consider the first two problems as eligible for conducting a replication paper as it studies whether the effectiveness generalizes for particular circumstances.       We consider problem-3 \textit{NOT} as eligible for conducting a replication paper as it does not validate the results of the original study but is an awareness measure independent of the video. 
\\\\
\textbf{Example Replications.}
We provide example replications (for which we assume that the addressed problem is eligible for conducting a replication paper) and their classification.

{Type-1}: The replication paper is based on the same domain, same method, and same analysis.

{Type-2}: The replication paper is based on the same domain, same method, but different analysis. There are various reasons why a different analysis is applied such as the assumptions of the statistical test are not met. 
%\textit{Only Data and Analysis are different}

{Type-3}: The replication paper is based on the same domain and same analysis. However, the method is different as the user study is conducted in the lab (and not as online survey anymore). 
%\textit{All Same but Method is different}
%5)	Andere Methode: Interview -> Survey; open text fragen; auch wieder open coden (Kleine methodische Änderung: survey -> lab)

{Type-4}: %(first example) 
The replication paper is based on the same domain. The method is different as the user study is conducted in the lab (and not as an online survey anymore). The analysis is also different. As in Type-2, there are various reasons why deciding to analyze the data in a different way.

{Type-5:} %(first example) 
The replication paper conducts the same survey using the same analysis while recruiting students in Switzerland, i.e. in a different domain.

{Type-6:} Using one of the domains from Type-5 but also using a different analysis as in Type-2.

{Type-7:} Using one of the domains from Type-5 but also using a different method as in Type-3.

{Type-8:} Using one of the domains from Type-5 but also using a different method and different analysis as in Type-4.

\section{RQ2: Systematic Literature Review}
%\textcolor{red}{welche Forschungsfrage(n) aus Intro werden hier beantwortet}

This section contributes to answer the following research question:
\textbf{RQ2:} \textit{\rqprevalence}
%\textbf{RQ3:} \textit{\rqreplicationtypes}

First, we explain our preconsiderations, followed by the literature search. 
\subsection{Preconsiderations}
\label{SearchConf}
%\textbf{.}
%\label{preconsiderations}
%\subsubsection{Searching for Replications}
\begin{comment}
    
We decided to search the following 13 venues \textit{CHI, SOUPS, USEC, EuroUSEC, CCS, CSCW, ICSE, PETS, S\&P, EuroS\&P, NDSS, WWW, and USENIX Security}:  we based our selection of venues on the analysis by Klemmer et al. \cite{klemmer_how_nodate}, which examined papers in the usable security and privacy field in terms of the availability of artifacts and details about their experiments. 
%It seems appropriate to adopt this selection of venues for our consideration, which represents the flip side or consequences of the availability of artifacts and information. %, namely the distribution of replication papers.
Note, in addition to their list of venues we searched in all EuroUSEC venues (and not just those being called symposium) as well as in the USEC publications. We also searched for publications at AsiaUSEC which only took place once in 2020. 
Furthermore, we decided to limit the search to the last 10 years, i.e., all publications between \textit{2016 and 2025}.
\end{comment}
While our goal was to identify those papers being a \textit{replication} according to the replication framework from Section~\ref{adoption}, we searched for word stems related to ``replicate'', ``reproduce'', and ``repeat''. Given that there are varying definitions of replication in the literature, we assumed authors would not consistently employ a single terminology. We chose these three because they were used by Olszewski et al. \cite{olszewski_sok_nodate} and 
Repeatability, % (Same team, same experimental setup)
%   The measurement can be obtained with stated precision by the same team using the same measurement procedure, the same measuring system, under the same operating conditions, in the same location on multiple trials. For computational experiments, this means that a researcher can reliably repeat her own computation.
Reproducibility, and % (Different team, different experimental setup )*
%    The measurement can be obtained with stated precision by a different team using the same measurement procedure, the same measuring system, under the same operating conditions, in the same or a different location on multiple trials. For computational experiments, this means that an independent group can obtain the same result using the author's own artifacts.
Replicability % (Different team, same experimental setup )*
%    The measurement can be obtained with stated precision by a different team, a different measuring system, in a different location on multiple trials. For computational experiments, this means that an independent group can obtain the same result using artifacts which they develop completely independently.
are the three terms used by ACM\footnote{\url{https://www.acm.org/publications/policies/artifact-review-badging} Last accessed: August 1st 2026}. %\textcolor{red}{stimmt die URL?} % for what can be considered replication studies.

Since the venues were published by different publishers, different approaches had to be taken depending on the publisher. Six of the 13 venues can be searched via the \textit{ACM Digital Library}\footnote{https://dl.acm.org/ Last accessed: August 1st 2026}, while two venues, S\&P and EuroS\&P, were searched via \textit{IEEE Xplore}\footnote{https://ieeexplore.ieee.org/Xplore/dynhome.jsp?tag=1 Last accessed: August 1st 2026}. The remaining venues (PETS, NDSS, USEC, USENIX Security, AsiaUSEC) were searched manually and with the help of scraping scripts. Depending on the years, EuroUSEC can be searched through the ACM Digital Library, IEEE Xplore, or manually. 
Table \ref{TabConf} shows which venues were searched by which method. For a more detailed description of each of the searches, including the exact search queries, see Appendix \ref{App:LitSearch}.

\begin{table}[!ht]
\scriptsize
    \centering
        \caption{* Search included workshops at the venue. \\$^\oplus$ "Security" and "Privacy" were added to the search string.}
        \begin{tabular}{|>{\centering\arraybackslash}p{0.25\linewidth}|>{\centering\arraybackslash}p{0.61\linewidth}|}
    \hline \rowcolor{gray!30}
        \textbf{Search Method} & \textbf{Venues} \\ \hline
        ACM Digital Library & \makecell{CHI*$^\oplus$, SOUPS, CCS*,ICSE*$^\oplus$, CSCW*$^\oplus$,\\ WWW*$^\oplus$, EuroUSEC (2021-2024)}\\ \hline
        \rowcolor{gray!10}
        IEEE Xplore & \cellcolor{gray!10}\makecell{S\&P, EuroS\&P*, EuroUSEC (2025)}\\ \hline
        \makecell{Manual and\\with own scripts} & \makecell{PETS, NDSS, USEC, USENIX Security,\\EuroUSEC (2016-2020), AsiaUSEC}  \\ \hline
    \end{tabular}

    \label{TabConf}
\end{table}

Furthermore, we decided to only consider user studies for our investigation. A \textit{user study} was defined as any empirical investigation that involves human participants directly. In particular, participants agreed that their data is used for research purpose. %. Participants can either be knowing or unknowing in case of deception studies.
%Studies that use user data that was not collected as part of the study, e.g. data from social media users, were not considered user studies. 
Studies that relied exclusively on user data collected outside of the study, such as social‑media posts, publicly available usage logs, or other secondary data, were not considered user studies.

\subsection{Description of PRISMA Diagram Steps}
The PRISMA diagram in Figure \ref{fig:PRISMA} provides an overview of the steps described below. 
\ \\\\
\begin{figure}[htbp]        
  \centering                   
  \includegraphics[width=0.51\textwidth]{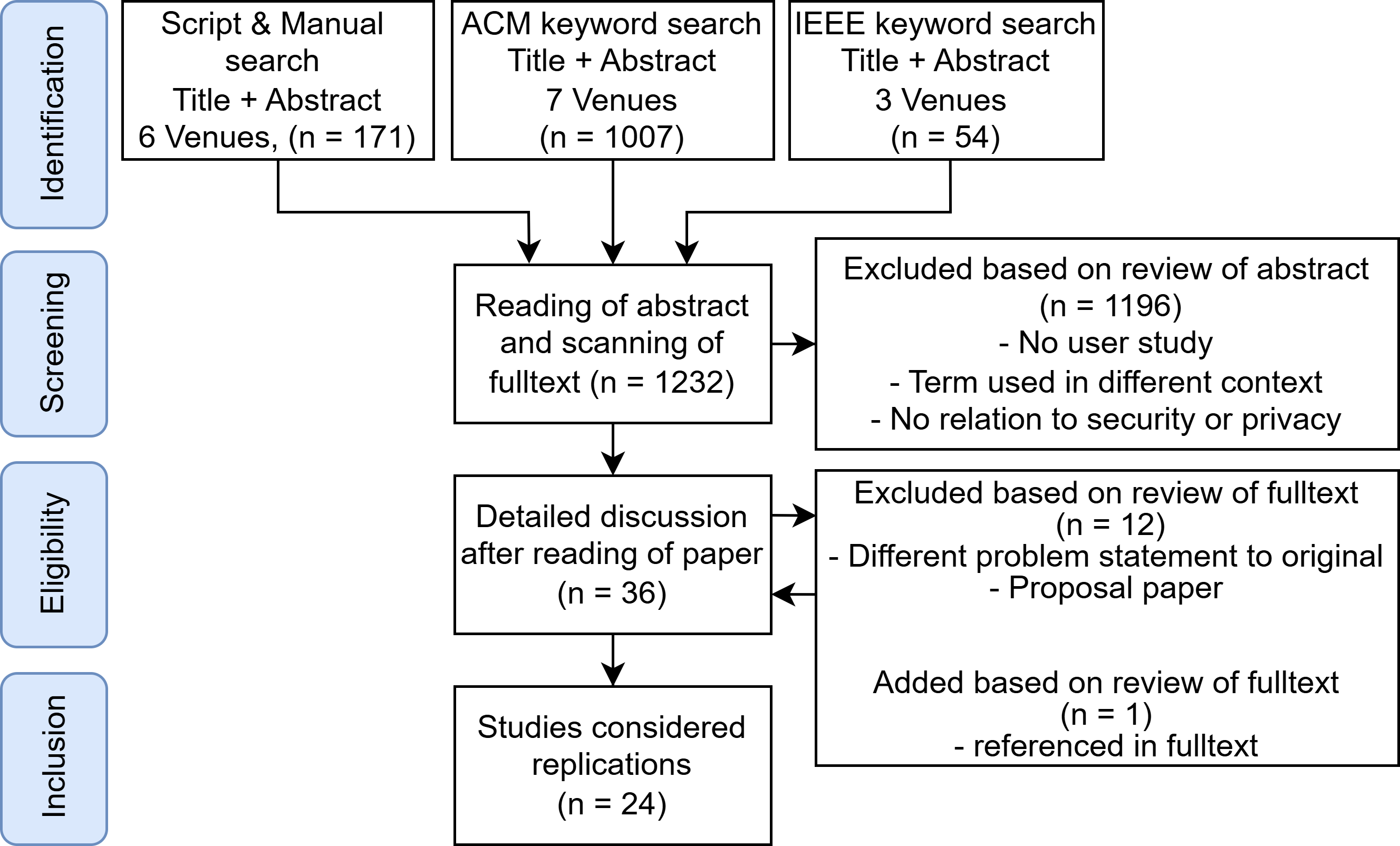}
  \caption{PRISMA diagram of the Literature Review. The six venues searched via script or manual include PETS, NDSS, USEC, USENIX Security, EuroUSEC (2016-2020) as well as the one AsiaUSEC. The three venues included in the IEEE search are S\&P, EuroS\&P and EuroUSEC (2025). The seven venues included in the ACM search are CHI, SOUPS, CCS, ICSE, CSCW, WWW and EuroUSEC (2021-2024).} %* The ACM search engine also included AI summaries containing the keywords when filtering for Title and Abstract, see Appendix \ref{App:LitSearch}
  \label{fig:PRISMA}
\end{figure}
\textbf{Identification.}
\label{serachQuerries}
%The following paragraphs provide a high level description for all 13 venues. 
First, we searched for papers that had either a variant of the word ``replicate'', ``reproduce'', or ``repeat'' in the title or abstract.  In addition, for CHI, ICSE, CSCW, and WWW we required the paper to either contain the term ``security'' or ``privacy'' since these venues address many topics beyond security and privacy. % in order to focus the search on papers dealing with privacy and security, these two terms (privacy, security) were added as prerequisites for the search, also in the context of the title and abstract. However, this additional restriction only applies to searches via ACM Digital Library and IEEE Xplore, which allow logical search queries. An exception is CCS, where this addition was omitted because all papers are related to security due to the nature of the venue.
When using ACM and IEEE for searching papers at the corresponding venues, the search included workshops for CHI\footnote{For CHI it was not workshops but late breaking research or alike. }, CCS, ICSE, CSCW, WWW, and EuroS\&P. We decided not to filter those out. % but instead checked for user study and replication as well.
When searching in the ACM Digital Library, we selected title and abstract while the search engine also searched in the AI-generated summary and the highlighting (see Appendix \ref{app:acm} for more details). The assignment of venues to the search variants can be found in Table \ref{TabConf}. EuroUSEC appears in all three search paths because the publisher has changed multiple times, which results in the number of venues in the PRISMA diagram not matching the number of venues in which we conducted our search.  %had not fixed it when finishing this submission. }. 

The initial search yielded a total of 1,232 papers. 
\ \\\\
\textbf{Screening.}
\label{manualInvest}
%The PRISMA flowchart in Figure \ref{fig:PRISMA} illustrates the systematic selection process that led to the final list of 24 replication papers.
Next, 1,196 publications were excluded after reviewing titles and abstracts, and where necessary, the full texts. Papers were removed if they lacked relevance to the field of usable security or privacy, if they did not involve a user study, and if they used the search terms in a context unrelated to replications. Note, in particular the search term ``repeat'' caused many false positives as it appeared but was used in a different context. Each paper was examined by two researchers before deciding to remove it.
\ \\\\
\textbf{Eligibility.}
\label{eligib}
Afterwards, the remaining 36 papers were read in detail. Two researchers examined each remaining paper whether they are replications according to our replication framework described in Section~\ref{adoption}, i.e., same problem / same research question. We noticed that even papers with replication in the title are no replication if we strictly require the authors to state that they have at least one research question in common with the original paper, i.e. their goal is to validate, question, or clarify their results. Therefore, we decided to include all papers when they actually validate, question, or clarify the results of the original paper. 
In Figure \ref{fig:PRISMA} we call these exclusion criteria a divergent problem statement compared to the original study. Note, while there is no restriction for the replication study being in the same paper as the original study, we do not consider the following cases as replications: (1) The goal of the paper is the development of some interventions which is iteratively developed, i.e. evaluated, improved, and re-evaluated using the same method. (2) Re-using a methodology but having an entire different goal (as in our example using the same methodology but testing different type of interventions) (3) running various experiments with slightly different methods as part of a scale development. Additional exclusion criteria at this stage was the paper being merely a proposal without actually having conducted the user study.

An additional replication paper was found while reading through all the papers in this step. This additional replication paper (\cite{naiakshina_if_2019}) was mentioned by Danilova et al. (2020) \cite{danilova2020replication} and was published at CHI 2019. The paper was not found through the keyword search as it does not contain any of the replication-related keywords (``replicate'', ``repeat'', ``reproduce'') in the title or abstract.  
\ \\\\
\textbf{Inclusion.}
\label{inclusion}
The result was a list of \textbf{24 papers}\footnote{for the complete list of replication papers go to \url{https://files.secuso.org/secuso/2026-08-Replication_Practices_Usable_Security_Privacy/} Last accessed: August 1st 2026} we identified as replications of user studies in the area of usable security and privacy.

\section{RQ3: Categorisation of Replications}
%\textcolor{red}{welche Forschungsfrage(n) aus Intro werden hier beantwortet}

This section contributes to answer the following research question:
%\textbf{RQ2:} \textit{\rqprevalence}
\textbf{RQ3:} \textit{\rqreplicationtypes}

\subsection{Methodology}
\label{CateorizationMethod}
In order to address the third research question, the 24 replication studies  were categorized into eight groups based on the framework described in Section~\ref{adoption}. 
To categorize the replication studies, they were distributed among three researchers, each of whom read the papers assigned to them. The findings were then jointly discussed and a decision was made for the three characteristics of domain, method, and analysis. The decision regarding the replication type assigned to each of the 24 papers is discussed in Section \ref{Categorize}. Note, in terms of being different, we did not distinguish between small and big changes. Any change was considered as doing it differently in this respect. Furthermore, in case there is more than one method applied in the original paper (e.g., quantitative and qualitative data being collected), we classified the replication already as different when one of the methods is different. The same holds for the other layers. %are different methods or different 
We discuss this approach in the discussion section.

%\subsection{Results}
\subsection{Some Numbers}
The majority of the replications were published at two venues: SOUPS (11) and CHI (6) (see Figure~\ref{fig:pieD}). %The relatively high number of replications at SOUPS can be explained by the explicit mention of replications in the call for papers for several years (see Subsection~\ref{CfP}). 
Also noteworthy is that of the original studies that were replicated by those 24 replications, 12 were published at SOUPS. We identified two papers from IEEE S\&P which got replicated. %It does not appear to have any effect on the venue of the replication study where the original study was published. 
%However, from the 298 SOUPS papers only 11 were replications.

\begin{figure}[htbp]        
  \centering                   
  \includegraphics[width=0.3\textwidth]{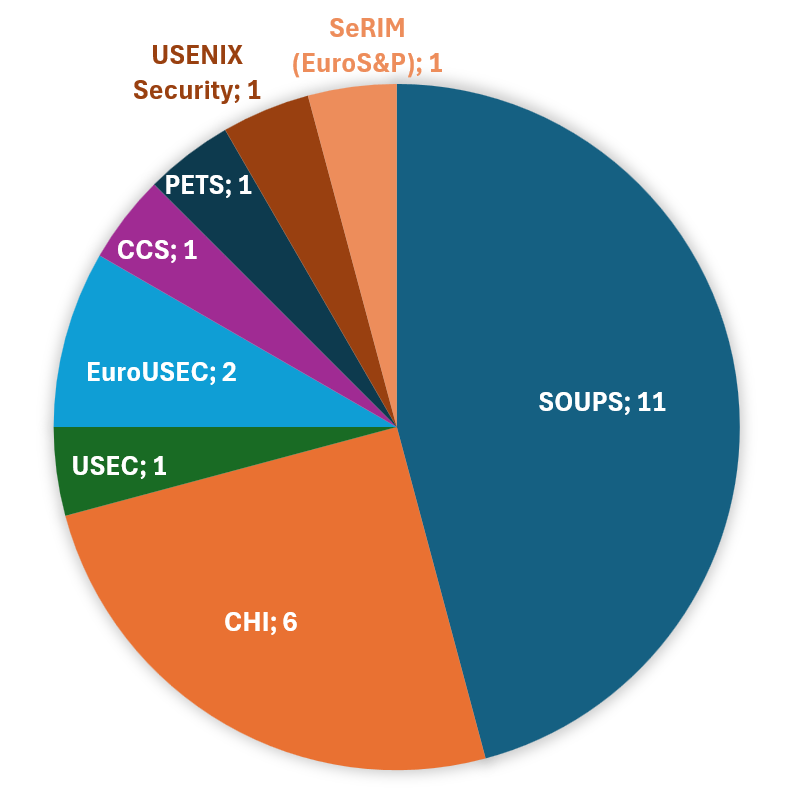}
  \caption{Distribution of replication studies per venue}
  \label{fig:pieD}
\end{figure}

%\begin{comment}

\begin{table*}[htbp]
\scriptsize
  \centering
  \caption{Background information about replication studies. \textbf{Term Used} lists the term used in the paper itself, about what it does. Papers that replicate more than one paper are represented with two rows, referring to both original papers. Column \textbf{Conference} lists the conference the replication paper was published at. If at least one author is part of the original as well as the replication paper, \textbf{Same Conference} lists if original paper and replication were published at the same conference. If ``No'', the conference at which the original paper was published is mentioned, \textbf{Same Author as Original} is stated as ``Yes''. \textbf{Years Between} shows how many years past between publishing of the original paper and the replication.}
  \label{tab:author-conf}
  \begin{tabular}{|l|p{2.3cm}|p{1.2cm}|p{1.7cm}|p{1.5cm}|p{2.5cm}|p{1cm}|}
    \hline \rowcolor{gray!30}
    \textbf{Replication Paper} &
    \textbf{Term Used} &
    \textbf{Conference} &
    \textbf{Original Paper} &
    \textbf{Same Author as Original} &
    \textbf{Same Conference} &
    \textbf{Years Between}\\
    \hline
        Naiakshina et al. (2019)~\cite{naiakshina_if_2019} &
    repeated &
    CHI &
    \cite{Naiakshina2017} &
    Yes &
    No (CCS) & 2\\
    
    \rowcolor{gray!10}Naiakshina et al. (2020) \cite{naiakshina2020conducting} &
     replicated the study&
    CHI &
    \cite{naiakshina_if_2019} &
     Yes &
     Yes & 1\\
    \rowcolor{gray!10}
     & follow-up Study 
     &
     &
    \cite{Naiakshina2017} &
     Yes &
    Yes & 3\\
    
    Mathis et al. (2021) \cite{mathis2021replicueauth} &
    conceptual replication &
    CHI &
    \cite{Khamis2018} &
    No &
    No (IMWUT)& 3\\
    
    \rowcolor{gray!10}Häring et al. (2023) \cite{haring2023less} &
    replication &
    CHI &
     \cite{haring21} &
    Yes &
    No (SOUPS)& 2\\
    
     Farzand et al. (2025) \cite{farzand2025scales} &
    conceptual replication &
    CHI &
    \cite{Bandara2020} &
    No &
    No (European Journal of Marketing) & 4\\
    
    \rowcolor{gray!10}Barber et al. (2025) \cite{barber2025beyond} &
    replication &
    CHI &
    \cite{Latulipe2022} &
    No &
    Yes & 3\\
    
    Busse et al. (2019) \cite{busse2019replicationSnooze} &
    extended replication &
    USEC &
    \cite{BravoLillo2013} &
    No &
    No (SOUPS) & 6\\
    
    \rowcolor{gray!10}Schessler et al. (2021)  \cite{schessler2021replication} &
    replication &
    EuroUSEC &
    \cite{Chin2012} &
    No &
    No (SOUPS) & 9\\
    
    Ortloff et al. (2021) \cite{ortloff2021replicating} &
    replication study &
    EuroUSEC &
    \cite{simoiu19} &
    No &
    No (SOUPS)& 2\\
    
    \rowcolor{gray!10}Canfield et al. (2017) \cite{canfield2017replication} &
    replication &
    SOUPS &
    \cite{Canfield2016} &
    Yes &
    No (Human Factors (Journal)) & 1\\
    
    Volkamer et al. (2018) \cite{volkamer2018replication} &
    replication study &
    SOUPS &
    \cite{DeLuca2010} &
    No &
    Yes & 8\\
    
    \rowcolor{gray!10}Al Qahtani et al. (2018) \cite{al2018effectiveness} &
    replication &
    SOUPS &
    \cite{Albayram17} &
    No &
    Yes & 1\\
    
    Hänsch et al. (2018) \cite{hansch2018programming} &
    partially replicate &
    SOUPS &
    \cite{Ceccato2013} &
    No &
    No (Empirical Software Engineering (Journal)) & 5\\
    
    \rowcolor{gray!10}Busse et al. (2019) \cite{busse2019replicationHack} &
    replication &
    SOUPS &
    \cite{Ion15} &
    No &
    Yes & 4\\
    
    Danilova et al. (2020) \cite{danilova2020replication} &
    replication &
    SOUPS &
    \cite{naiakshina_if_2019} &
    Yes &
    No (CHI) & 1\\
    
    \rowcolor{gray!10}Baig et al (2021) \cite{baig2021replication} &
    replication &
    SOUPS &
    \cite{Fulton19} &
    No &
    Yes & 2\\
    
    Tang et al. (2022) \cite{tang2022replication} &
    replication &
    SOUPS &
    \cite{redmiles19} &
    No &
    No (IEEE S\&P) & 3\\
    
    \rowcolor{gray!10}Kühtreiber et al. (2022) \cite{kuhtreiber2022replication} &
    replication &
    SOUPS &
    \cite{Xiong2020} &
    No &
    No (IEEE S\&P) & 2\\
    
    Pfeffer et al (2022) \cite{pfeffer2022replication} &
    replication &
    SOUPS &
    \cite{Rader2012} &
    Yes &
    Yes & 10\\
    
    \rowcolor{gray!10}Ortloff et al. (2025) \cite{ortloff2025replication} &
     re-replication &
     SOUPS &
    \cite{Ion15} &
     No &
     Yes & 10\\
     \rowcolor{gray!10}&
     re-replication &
     SOUPS&
    \cite{busse2019replicationHack} &
     Yes &
     Yes & 6\\
    
    Warberg et al. (2019)\cite{warberg2019can} &
    replication &
    CCS &
    Replicating itself &
    Yes &
    Yes & 0*\\
    
    \rowcolor{gray!10}Ismail et al. (2017) \cite{ismail2017permit} &
    large scale replication &
    PETS &
    \cite{Ismail15}&
    Yes &
    No (CHI) & 2\\
    
    Noah et al. (2025) \cite{noah2025replication} &
    replication study &
    \makecell{SeRim\\(EuroS\&P)} &
    \cite{duzgun22} &
    Yes &
    No (NordiCHI) & 3\\
    
    \rowcolor{gray!10}Ray et al (2021) \cite{ray2021older} &
    replication  &
    USENIX Security &
    \cite{Pearman19} &
    No &
    No (SOUPS) & 2\\
    \hline

  \end{tabular}
\end{table*}

All but one paper, called there research replication while some use extensions such as partial replication, extended replication and re-replication. The one paper which does not use the term replication used the term repeated. 
Eight out of the 24 replication papers included at least one author from the original study. One of the papers replicated two different papers and one paper is a self-replication. 

Furthermore, five papers replicated the research within one year of the original publication, and seven papers did so within two years. %(3 with overlapped authorship)
In two cases, the longest time interval was identified as 10 years between original publication and replication. One replication from different authors was published one year after the original one. %within a one year time frame between publications.
Examining the demographic composition of the author affiliation country, 14 papers originated from Germany (10 exclusively, 4 in collaboration) and 9 from the U.S. (5 exclusively, 4 in collaboration). %, 4 from the United Kingdom (2 exclusively, 2 in collaboration), and one each from Canada (exclusively), Finland, Denmark, Qatar, Saudi Arabia, Austria and Switzerland (all in collaboration). 
Notably, nine of the German publications had the same institution (University of Bonn) among the authors while eight had Matthew Smith as co-author. 
Eight of the 11 SOUPS papers, as well as two of the three EuroUSEC papers and the one USEC paper used the term ``replication'' in the title as required by the CfP\footnote{For more information about these 24 papers see \url{https://files.secuso.org/secuso/2026-08-Replication_Practices_Usable_Security_Privacy/} Last accessed: August 1st 2026}.

\subsection{Categorization of Replication Studies}
\label{Categorize}

Figure~\ref{validityTree2} provides an overview, where we assigned the papers in the adapted replication framework~\cite{olszewski_sok_nodate}. None of the replications was classified as exact replication (type 1 / type 2). None of the papers was classified as type 6 (which would be different domain, same method but different analyses). 19 replicated in a different domain, 19 replicated using a different method (14 replicated in both a different domain and using a different method), and 7 replicated using a different analysis (all 7 used a different method, too). 

\

For each of the 24 papers, we very briefly summarize the common problem addressed in both the original and the replication paper; and mention those layers of the framework (domain, method, analysis) which are different by highlighting the differences or at least one of the differences. 

\ \\\textbf{Two replication papers assigned to type-3}, i.e. same domain and same analysis but different method.

% 15. \cite{danilova2020replication}
%org \cite{naiakshina_if_2019}
Danilova et al. (2020)~\cite{danilova2020replication} replicates the developers password storage task study from Naiakshina et al. (2019)~\cite{naiakshina_if_2019} but without deception (method different). 
%(Type-3)

% 21. \cite{warberg2019can}
An unusual case presents itself in Warberg et al. (2019)~\cite{warberg2019can}, a paper investigating psychometrically tailored nudges of data disclosure choices. A replication occurs within the same paper. Instead of repeating a study to improve scales, questionnaires, etc., a study is repeated to examine the robustness of the results. Changes to the structure of the study resulted in removing a follow up survey and therefore a change in the method. %(Type-3)%, the study stays the same for domain and analysis. (Type-3)

\ \\\textbf{Three replication papers assigned to type-4}, i.e. same domain but different method and different analysis.

% 4. \cite{haring2023less}
%original \cite{haring21}
A replication paper about the German COVID-19 contact tracing app by Häring et al. (2023)~\cite{haring2023less} compares opinion about the app after its release with opinions before the app release~\cite{haring21}. The phrasing of the survey questions were adapted and thus the method changed. For the replication the authors decided to not only code their own qualitative data but also re-coding the qualitative data from the original study with a codebook developed by another study about COVID-19 apps (different analysis). 

% 17. \cite{tang2022replication}
%org \cite{redmiles19}
Tang et al. (2022)~\cite{tang2022replication} replicates the study from~\cite{redmiles19} regarding the external validity of privacy and security surveys conducted on MTurk. The survey questions were changed (method) and the groups are not compared to each other just to the control group from Pew (analysis).

% 7. \cite{busse2019replicationSnooze}
%org \cite{BravoLillo2013}
The replication paper by Busse et al. (2019)~\cite{busse2019replicationSnooze}, which focuses on the habituation effect for security warnings, adds a monetary incentive to the study apart from other smaller changes, changing the method compared to the original study \cite{BravoLillo2013}. %Due to other changes to the study set up, described in detail through an overview table in the study, 
The analysis also differs: different test with chi-square; pointing to not being able to replicate original data. 

\ \\\textbf{Five replication papers assigned to type-5}, i.e. different domain but same method and same analysis.

The replication paper by Canfield et al. (2017)~\cite{canfield2017replication} only replicates part of their original study~\cite{Canfield2016} about phishing and changed the domain by recruiting participants locally instead of MTurk resulting in a participant pool consisting of mainly retirees and college students. %(Type-5)

% 14. \cite{busse2019replicationHack}
%org \cite{Ion15}
Busse et al. (2019)~\cite{busse2019replicationHack} replicates the investigation of~\cite{Ion15} in self-reported security behavior of security experts and non-experts. The sample of one part of the replication is different as they recruited European security experts in contrast to the mostly U.S. sample from the original work (i.e., the domain is different). % (Type-5)
% 1.\cite{naiakshina_if_2019}

% 16. \cite{baig2021replication}
%org \cite{Fulton19}
Baig et al. (2021)~\cite{baig2021replication} replicates an investigation about the effect of media on people's mental models of security~\cite{Fulton19}. Unlike the original study, Baig et al. (2021) focuses on technical users (i.e., different domain). %(Type-5)

Kühtreiber et al. (2022)~\cite{kuhtreiber2022replication} studies the effect of differential privacy communication on German users’ comprehension and data sharing attitudes which was studied in~\cite{Xiong2020} for users from the U.S. and India (different domain). Note, the original paper describes four studies, the replication paper only two. % (Type-5)%For those experiments being replicated, the methods and the analysis were the same. (Type-5)

% 24. \cite{ray2021older}
%org \cite{Pearman19}
Ray et al. (2021)~\cite{ray2021older} replicates a study investigating password mangers adoption~\cite{Pearman19}. The authors only changed the domain by focusing on older adults compared to a predominantly younger participant sample. For the analysis they used the same codebook the original study used.% (Type-5)

\begin{figure}[htbp]
    \centering
    \resizebox{0.49\textwidth}{!}{
    \begin{tikzpicture}[
    node/.style={circle, draw=gray, fill=white, thick, minimum size=6mm},
    same/.style={->, very thick, blue!70!black},
    diff/.style={->, very thick, orange!80!black},
    repli/.style={rectangle, draw=green!60, fill=green!5, thick},
    colline/.style={gray!60, dashed, thick},
    >=Stealth
]

% Column x-positions
\def\xP{-0.5}
\def\xD{2}
\def\xM{4.5}
\def\xA{7}

% Column titles
\node at (\xP+0.375,5.2) {\Large{\textbf{Problem}}};
\node at (\xD,5.2) {\Large{\textbf{Domain}}};
\node at (\xM,5.2) {\Large{\textbf{Method}}};
\node at (\xA,5.2) {\Large{\textbf{Analysis}}};

% Colored Spaces
\fill[gray!25] (-1,2.5) rectangle (12.2,4.8);
\fill[gray!5] (-1,-4.6) rectangle (12.2,2.6);

% Vertical separators
\draw[colline] (\xD-1.25,-4.7) -- (\xD-1.25,5);
\draw[colline] (\xM-1.25,-4.7) -- (\xM-1.25,5);
\draw[colline] (\xA-1.25,-4.7) -- (\xA-1.25,5);

% Nodes
\node[node] (P)  at (\xP+0.375,0) {};

\node[node] (D1) at (\xD, 2.5) {};
\node[node] (D2) at (\xD,-2.5) {};

\node[node] (M11) at (\xM, 3.6) {};
\node[node] (M12) at (\xM, 1.2) {};
\node[node] (M21) at (\xM,-1.2) {};
\node[node] (M22) at (\xM,-3.6) {};

% Analysis
\node[node] (A1) at (\xA, 4.2) {1};
\node[node] (A2) at (\xA, 3.0) {2};
\node[node] (A3) at (\xA, 1.8) {3};
\node[node] (A4) at (\xA, 0.6) {4};

% lower

\node[node] (A5) at (\xA,-0.6) {5};
\node[node] (A6) at (\xA,-1.8) {6};
\node[node] (A7) at (\xA,-3.0) {7};
\node[node] (A8) at (\xA,-4.2) {8};

% Labels on analysis nodes
\node[right=3mm] at (A1) {};
\node[right=3mm] at (A2) {};
\node[right=3mm] at (A3) {\large{\cite{danilova2020replication,warberg2019can}}};
\node[right=3mm] at (A4) {\large{\cite{busse2019replicationSnooze,haring2023less,tang2022replication}}};
\node[right=3mm] at (A5) {\large{\cite{canfield2017replication,busse2019replicationHack,baig2021replication,kuhtreiber2022replication,ray2021older}}};
\node[right=3mm] at (A6) {};
\node[right=3mm, text width=5cm] at (A7) {\large{\cite{mathis2021replicueauth,farzand2025scales,barber2025beyond,ortloff2021replicating,volkamer2018replication,noah2025replication,naiakshina2020conducting,pfeffer2022replication,ortloff2025replication,naiakshina_if_2019}}};
\node[right=3mm] at (A8) {\large{\cite{schessler2021replication,ismail2017permit,al2018effectiveness,hansch2018programming}}};

% Edges: Problem -> Domain
\draw[same] (P) -- (D1);
\draw[diff] (P) -- (D2);

% Domain -> Method
\draw[same] (D1) -- (M11);
\draw[diff] (D1) -- (M12);
\draw[same] (D2) -- (M21);
\draw[diff] (D2) -- (M22);

% Method -> Data
\draw[same] (M11) -- (A1);
\draw[diff] (M11) -- (A2);
\draw[same] (M12) -- (A3);
\draw[diff] (M12) -- (A4);
\draw[same] (M21) -- (A5);
\draw[diff] (M21) -- (A6);
\draw[same] (M22) -- (A7);
\draw[diff] (M22) -- (A8);

% Legend
\draw[same] (2.2,-4.9) -- +(1,0) node[right, black] {\large{Same}};
\draw[diff] (4.8,-4.9) -- +(1,0) node[right, black] {\large{Different}};

\filldraw[black] (-0.8,3.6) circle (0pt) node[anchor=west]{\Large{Exact Replication}};
\filldraw[black] (-0.8,-3.8) circle (0pt) node[anchor=west]{\Large{Conceptual Replication}};

\end{tikzpicture}
    }
    \caption{Categorization of replication paper}%\textcolor%{green}{bekommen wir das auf eine spalte?
  %  }} % through \textit{Tree of Validity} \cite{olszewski_sok_nodate}}
    \label{validityTree2}
\end{figure}
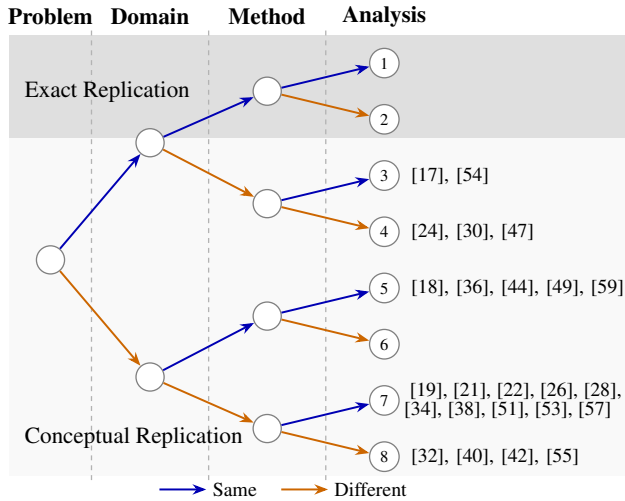

\ \\\textbf{Ten replication papers assigned to type-7}, i.e. different domain and different method but same analysis.

Naiakshina et al. (2019)~\cite{naiakshina_if_2019} replicated their developer password storage study~\cite{Naiakshina2017} by shifting the domain from undergraduate computer science students to freelance programmers, and changing the method by adding a deception element.% (Type-7)

% 2. \cite{naiakshina2020conducting}
% original: \cite{naiakshina_if_2019}
The replication paper by Naiakshina et al. (2020)~\cite{naiakshina2020conducting} replicates a study by Naiakshina et al. (2019)~\cite{naiakshina_if_2019} which is also a replication study of Naiakshina et al. (2017)~\cite{Naiakshina2017}. %The study investigates the approach taken by developers when storing passwords. 
Naiakshina et al. (2020) changes domain by considering developers who are employed by German companies, whereas Naiakshina (2019) considered freelance developers and Naiakshina et al. (2017) studied students. The method for all three is not the same, as already the 2017 and 2019 paper have different methods in place (due to one using deceptive elements).

% 3. \cite{mathis2021replicueauth}
% original \cite{Khamis2018}
Mathis et al. (2021)~\cite{mathis2021replicueauth} investigates whether the use of VR for conducting a study remotely is viable as a substitute for in-person laboratory or field studies. To do so, it replicates an authentication field study~\cite{Khamis2018} by changing the method (authentication setup was in VR) of the first part and the domain in the second part of the study. The second part of the study was conducted remote for the replication, while the original was carried out in a laboratory setting, changing the recruitment process. % and as a side effect also delivers comparable results who validate the original results. 
%(Type-7)

% 5. The paper 'When Scales Fail to Measure Up: How Not to Measure Social Media Privacy–Findings of a Representative Survey in 16 Countries'
% original \cite{Bandara2020}
Farzand et al. (2025)~\cite{farzand2025scales} validates a privacy scale in 16 Middle East and North Africa countries, which the original paper~\cite{Bandara2020} introduced and validated in Australia. Thus, the domain is different. Their methodology is slightly different as they changed from a 5-point to a 7-point likert scale for several questions. %Regarding the anaylses being the same or different, we conclude that it is the same: 
For their analysis, they checked whether their data fits to the model %while the original They checked whether their data fits to the model 
while the original paper developed the model based on their data, which we consider to be the same analysis. % (Type-7)

% 6. \cite{barber2025beyond}
% orignal \cite{Latulipe2022}
Barber et al. (2025)~\cite{barber2025beyond} replicates a study about digital banking adoption by older adults and how ``close others'' like relatives help them. They change the domain from Canada to the U.K. and include the older adults in the survey, whereas the original only surveys ``close others''~\cite{Latulipe2022}. It is difficult to assess the extent to which the questionnaire itself has changed, as neither the replication nor the original paper include the questionnaire. The replication did analyze their data with a set of statistical tests not present in the original study. %(Type-7)

% 9. \cite{ortloff2021replicating}
%org \cite{simoiu19}
The paper by Ortloff et al. (2021)~\cite{ortloff2021replicating} aims to replicate a study~\cite{simoiu19} about ransomware in the US but for Germany. Apart from the domain, the method was changed slightly by adding more detailed definitions of ransomware. %(Type-7)%The resulting data and analysis was consistent to the original study. 

% 11. \cite{volkamer2018replication}
% org \cite{DeLuca2010}
The replication paper by Volkamer et al. (2018)~\cite{volkamer2018replication} replicates the study described in~\cite{DeLuca2010} which studied whether and why people try to hide when entering a PIN at an ATM. While the original paper observed and interviewed people in Germany and The Netherlands, the replication was conducted in Germany, U.K., and Sweden (i.e., the domain is different). The method was slightly adapted by ‘incorporate design aspects from their lessons learned section'~\cite{volkamer2018replication} of the original paper.% (Type-7)

% 19. \cite{pfeffer2022replication}
% org \cite{Rader2012}
Pfeffer et al. (2022)~\cite{pfeffer2022replication} conductes a repliaction paper regarding anecdotal stories about security threats and their effect on people with slight changes to the original study~\cite{Rader2012}. They change the domain by using a divers sample of participants instead of only students. The survey was adapted  to be more contemporary and adding one additional question at the end of the questionnaire to avoid priming. %(Type-7) %Due to the only minimal adjustments to the questions the method can be considered to be the same to the original. (Type-5) \textcolor{red}{CM: Würdet ihr bei method = same hier zustimmen? MV: ich würde jede änderung ausser übersetzung als geänderte methode sehen}% und später in discussion diskutieren ob das sinnvoll ist // BB: Würde auch sagen, dass man einfach hart bleibt und alles was nicht maximal übersetzt ist, einfach anders ist -> dann disktutieren, dass anhand der Daten im Paper auch oft schwer einzuschätzen ist wie ähnlich es ist}

%\begin{takeaways} hatten wir schon
 %   Changes to the questionnaire were mentioned and the questionnaire is displayed in the Appendix. But the exact changes are not clearly indicated. 
%\end{takeaways}

% 20. \cite{ortloff_qualitative_2025}
% org \cite{Ion15}
The paper by Ortloff et al. (2025)~\cite{ortloff2025replication} is a replication of Busse et al. (2019)~\cite{busse2019replicationHack} which again is a replication of~\cite{Ion15}. 
The original paper \cite{busse2019replicationHack} consists of interviews and survey with two different groups (security experts/non-experts). The domain changes (use of Prolific instead of MTurk and they aimed for a representative sample) as well as the method (during the interview they showed results of the original studies). 

% 23. \cite{noah2025replication}
%org \cite{duzgun22}
Noah et al. (2025)~\cite{noah2025replication} replicates a user study on recognition-based graphical authentication in AR. While the original paper~\cite{duzgun22} conducted the study in Germany, Noah et al. (2025) did so in the U.S. (changing the domain). They also changed the method as the implementation of the authentication scheme differs, as seen on the images shown in the papers. %slightly but 

\ \\\textbf{Four replication papers assigned to type-8}, i.e. the replication differs in all three layers.

% 8. \cite{schessler2021replication}
% org \cite{Chin2012}
In a study about the differences in security perception between smartphone and laptop users Schessler et al. (2021)~\cite{schessler2021replication} replicates the original study by Chin et al. (2012)~\cite{Chin2012}. They changed the country from the US to Germany. Due to COVID-19 regulations the interviews took place online and content-wise changes were made leading to a change in method. The analysis differs due to changes in the selection of some statistical tests and adjustments to scales. %(Type-8)

% 12. \cite{al2018effectiveness}
% org \cite{Albayram17}
Al Qahtani et al. (2018)~\cite{al2018effectiveness} replicates Albayrami et al. (2017)~\cite{Albayram17} which examines fear appeals through videos in connection with smartphone security risks in Saudi Arabia. %, differs from the original study \cite{} in all four categories. 
It takes place in another country (domain), instead of open-ended questions multiple choice questioned were employed based on the codebook from the original study (method) and no own codebook was developed. As Al Qahtani et al. (2018) do not go into detail about their post-hoc statistical test we can not confirm, that the analysis was done with the same statistical tests (analysis).% (Type-8) 

% 13. \cite{hansch2018programming}
%org \cite{Ceccato2013}
Hänsch et al. (2018)~\cite{hansch2018programming} revisits the study by Ceccato et al. (2013)~\cite{Ceccato2013} which investigates the effect on software obfuscation techniques for reverse engineering protection. Participants are students from Germany instead of the UK and Italy, changing the domain. The method by Hänsch et al. (2018) is slightly different as they omit questions in the survey. Hänsch et al. (2018) used a different non-parametric analysis for the within factors and therefore the analysis is different.

% 22. \cite{ismail2017permit}
% org \cite{Ismail15}
Ismail et al. (2017)~\cite{ismail2017permit} replicates an investigation in how permission options affect usability for social media apps~\cite{Ismail15}. The domain switched  from students to recruiting through MTurk. They change the method and analysis by changing scales (SEQ and SUS instead of self-developed ones) and changing from a within-subjects experimental design to a between-subjects design. %(Type-8)

\subsection{Noteworthy Insights}
\label{Noteworthy_Insights}
When identifying, analyzing, and categorizing the papers, we identified noteworthy aspects of replication papers\footnote{Note, we mention at least one paper in which we noticed the particular aspect. However, it might be that the aspect holds for other papers, too.}, i.e. aspects worth considering when later on deducing recommendations for venues, authors, and reviewers.  %was noticed in a gray box the first time they appeared, independently of whether they were mentioned later in other papers.

\ \\\textbf{General exclusion criteria.} In general, only a few papers explicitly stated that their research question was the same and / or they wanted to validate the original findings. Therefore, we included all those papers which did so implicitly, e.g. by comparing results to the original studies.
%We excluded several papers from being a replication as they 'only' reused (parts of) the methodology (but different research question), \textcolor{red}{we kept~\cite{mathis2021replicueauth} as we considered it as edge case}. 
Work that conducts the same study in different countries is also not classified as a replication within a single paper, and we do not take it into account.

\ \\\textbf{Some difficulties when categorizing papers.} For some replication papers, it was challenging to categorize them due to missing information such as~\cite{barber2025beyond,hansch2018programming}, and~\cite{pfeffer2022replication} while other papers such as~\cite{busse2019replicationSnooze} and~\cite{danilova2020replication} made it easy by providing tables explaining the differences. In general, it was difficult to categorize those papers being a replication of a paper which is itself a replication (e.g. the chain from ~\cite{naiakshina2020conducting} to  ~\cite{naiakshina_if_2019} to~\cite{Naiakshina2017}). The same holds if the original paper conducted more than one study, e.g.~\cite{busse2019replicationHack,tang2022replication}.

\ \\\textbf{Differences in domain.} While there are many ways to change the domain, we find the one from \cite{ortloff2025replication} worth mentioning as they changed to a representative sample. 
11 papers changed the domain by conducting the study in a different country (e.g.~\cite{al2018effectiveness,schessler2021replication,ortloff2021replicating}). This comes with the challenge of ensuring the translation is true to the original, discussed in~%Translation issue when replicating in a different country
\cite{ortloff2021replicating}.

\ \\\textbf{Differences in method.}
There are different ways to change the method. Interestingly, the authors of~\cite{naiakshina_if_2019}  change from a deceptive study to a non deceptive one. Furthermore, we found in ~\cite{haring2023less} a change in questions  due to change in situation (intention to use was studied in the original paper versus actual usage in the replication paper).  We also noticed that there are in some cases very small changes, such as switching from a 5‑ to a 7‑point Likert scale~\cite{farzand2025scales} which we considered as difference as the impact is unclear. 
Changes to data collection are necessary in cases where data protection regulations contain requirements that are not met by the original study, such as making responses to certain questions optional.

\ \\\textbf{Differences in analysis.}
We noticed from~\cite{farzand2025scales} that depending on the quantitative analyses applied in the original paper, ``same'' can have different interpretations; e.g. if the original paper used open coding, should the replication do so as well, or use the code book from the original paper to apply closed coding, as in~\cite{ray2021older}.

\ \\\textbf{Issues with original study.}
We identified several cases in which issues with the original study were mentioned as reasons for adoptions when replicating the research: The authors of~\cite{haring2023less} re-coded the data from the original paper due to issues they saw in the original paper. The authors of \cite{busse2019replicationHack} noticed some issues with some questions. In order to get comparable results, they implemented two groups (one with the original questions and one with their modified questions).  

The authors of~\cite{volkamer2018replication} incorporated lessons learned from original paper on the studied method. 
The authors of~\cite{noah2025replication} adopted the study authentication scheme due to some critique on design decisions of the original paper. 
Some replication studies such as~\cite{ismail2017permit} move away from the self developed scales from the original study to more established scales.
Note, this approach is inline with the statements in the SOUPS CfP: ``[..] use currently accepted methodologies and technologies. Authors should not reuse outdated methods / technologies simply because they were used in the original paper.''

\ \\ \textbf{Replication versus replication-extension versus partial replication. }
Several papers replicated a paper but also answered additional research questions, such as~\cite{ortloff2025replication, barber2025beyond,baig2021replication}. The authors of~\cite{barber2025beyond} even call it a ``replication-extension''. The authors of~\cite{baig2021replication} explained they put the additional questions at the end to reduce or mitigate influence on replicated data. 
There was also one replication paper~\cite{canfield2017replication} which can be considered as partial replication, as only one of the two studies from the original paper was replicated. 

\ \\\textcolor{black}{\textbf{Section structure of replication papers.} There is neither a pattern on how the paper is structured nor how the different sections are structured. Furthermore, we kind of assumed that the related work sections of replication papers are similarly structured as the corresponding section in the original paper; and provide an extension of the related work by newer publications. %based on the related work cited in the original paper, nor do
We also assumed to see related work on replication papers, which is not the case for the 24 papers. We also see different approaches in the ``Methodology'' section. While some studies barely refer to the original methodology (such as \cite{ismail2017permit}), others %most do so either by referencing the original methodology without explaining the details of changes (such \cite{schessler2021replication}) or through 
devoted subsections to the changes compared to the original study (such as \cite{naiakshina2020conducting, danilova2020replication}). Differences are sometimes also presented in tabular form (e.g., in \cite{busse2019replicationSnooze}). Regarding the result section, we made the following observations: In most papers, the results of the original study are presented directly in the ``Results'' section alongside the newly collected data (such as \cite{danilova2020replication,mathis2021replicueauth, ray2021older}). A few papers included the comparison to the original paper only in the ``Discussion'' section or barely at all (such as \cite{al2018effectiveness,ismail2017permit}}).

\section{RQ4: Survey with Authors}
In this section, we want to answer \textbf{RQ4:} \textit{\rqsurvey{}}

\begin{table*}[htbp]
\footnotesize
  \centering
  \caption{Themes identified with the codes with a short description and frequency of being named by the participants.}
  \label{tab:freq}
  \begin{tabular}{|l|p{9.5cm}|r|}
  \hline \rowcolor{gray!30}
    \textbf{Code} & \textbf{Definition} & \textbf{Frequency} \\
    \hline
    Scientific Validation \& Generalizability       & To confirm, extend, or generalize earlier results to new populations, settings, or theoretical frameworks & 26 \\
    \rowcolor{gray!10}Temporal Evolution \& Contextual Change         & Evolving technologies, shifting threat models, or new contextual factors prompt replication to test the durability of findings over time & 19\\
    Operational Efficiency \& Resource Availability & The ease of access to artifacts, the simplicity of the protocol, and savings in effort or cost shape the decision to replicate & 19\\
    \rowcolor{gray!10}Author Engagement \& Intent                     & The relationship to the original work (e.g., being the original author), the planned versus opportunistic nature of the replication, and pedagogical motives influence the decision & 11 \\
    Trust \& Reliability Concerns                   & Doubts about the credibility of the published results or the rigor of the original methodology trigger a replication effort & 8 \\
     \rowcolor{gray!10}Replication Barriers \& Constraints             & Practical limitations—such as resource shortages, methodological hurdles, or technological drift—often prevent a replication from being attempted & 6 \\
    Personal Motivation \& Advocacy                 & Replication decisions are driven by personal enthusiasm or a commitment to promote replication as a scientific norm & 5 \\
    \rowcolor{gray!10}Special themes                                  & Some smaller themes found that did not fit other categories & 3\\
    \hline
  \end{tabular}
\end{table*}

\subsection{Methodology}
\
%\textbf{Procedure.} 
We emailed the authors of the 24 replication papers identified in our systematic review, while we did not send emails to ourselves. %, naming each study explicitly. 
Each author was asked to submit one response, even if they appeared on multiple papers, so the motivations would reflect their individual perspective. We encouraged independent replies from multiple authors of the same papers to capture varying viewpoints within the same team. 17 emails could not be delivered as the account was not available anymore.
25 authors participated. % completed the questionnaire.
\ \\\\
\textbf{Questionnaire.} We designed the survey explicitly with the busy schedules of researchers in mind and therefore kept it as short as possible while still being able to answer our research question. It consisted of an informed consent and once agreed there were only four questions (where the last one was worded differently based on the response to the third one). No sociodemographic or identifying information (e.g. institutional affiliation, email address) was collected. The complete questionnaire can be found in Appendix \ref{appendix:authorsurvey}.
\ \\\\
\textbf{Analysis.} We used an inductive, two‑phase coding protocol that followed the standard practices of usability‑oriented qualitative research~\cite{richards2018practical}. 

\label{sec:data-prep}

\label{sec:open-coding}
Two researchers independently read every answer and applied codes to statements regarding reasons to conduct a replication paper. No predefined taxonomy was imposed; each code was created de‑novo to capture the participants expressed motivation or rationale. Discrepancies in the coding were discussed among the researchers after a first round of coding. %The initial code counts were R1: 115 codes, R2: 90 codes.  
For all codes exemplary quotations were collected during the first round to minimize ambiguity during discussions.
%\textbf{Code Reconciliation and Merging.}
\label{sec:merge}
%After the independent coding phase, the two coders performed a structured comparison session using MaxQDA.  %The similarity matrix highlighted overlapping codes, which were merged according to the criteria in Table~\ref{tab:merge-criteria}.
% \begin{table}[htbp]
%   \centering
%   \caption{Criteria for merging codes}
%   \label{tab:merge-criteria}
%   \begin{tabular}{p{0.15\linewidth}p{0.75\linewidth}}
%     \toprule
%     \textbf{Criterion} & \textbf{Description} \\
%     \midrule
%     Semantic overlap & Both codes refer to the same underlying construct (e.g.\ ``Confirm Result'' vs.\ ``Validation, Robustness, Scientific Rigor''). \\
%     Lexical similarity & Minor spelling or grammatical differences (e.g.\ ``Easier to do'' vs.\ ``Easy to do''). \\
%     \bottomrule
%   \end{tabular}
% \end{table}
%\subsection{Result}
%The reconciliation produced 39 shared codes (75\% of the original set). The remaining 13 codes were retained as unique because they represented distinct motivations (e.g., ``Material was available'') or lacked a counterpart in the other coder’s list.  Sub‑code hierarchies were introduced for broad codes. % (e.g.\ ``Temporal Evolution \& Contextual Change'' $\rightarrow$ ``ChangesOverTime'', ``TechnologyDevelopment''). 
%\textbf{Intercoder Reliability Assessment.}
\label{sec:reliability}
In the second phase, both coders re‑coded the data independently. The intercoder agreement reached $85.3\%$, indicating %, surpassing the 0.80 threshold. %commonly cited for 
acceptable reliability \cite{krippendorff2004reliability}.

%\textbf{Final Codebook and Thematic Mapping.}
\subsection{Results}
\label{sec:final-codebook}
All answers were coded together. We identified 42 codes (to view the full codebook feel free to contact the authors from the Karlsruhe Institute of Technology). These were categorized in general reasons to replicate (or not to replicate) and characteristics of studies for which the motivation to replicate is particularly high. For the first category (reasons to replicate), eight themes were identified from 29 codes (see Table~\ref{tab:freq} for these themes). The most prevalent reason to conduct a replication ($N=26$) was to validate or generalize the results % from a previous study 
(e.g., \textit{``We wanted to know to what extent our original results would replicate in a sample with different demographics [...].'' -P3}).

Nearly as often ($N=19$), researchers mentioned either the higher efficiency due to available material or methods  (e.g.,  \textit{``We also chose to replicate this specific user study because study material was available to us  [...]''} -P21) or the expected changes over time from a previous study as reasons to replicate it (\textit{``To see if recent developments in tech / IT security / politics have had an impact on the study subject.''} -P23).
The least mentioned reasons ($N=5$) to replicate included "Personal Motivation \& Advocacy"
   (e.g., \textit{ ``[...], i chose to expand on the topic bc i am personally a big fan of replication.''} -P11).

%
%For the question ``What studies should be replicated''  we collected the following answers and their frequency (shown in the appendix in Table~\ref{tab:whatfreq}). 
For the second category, we decided to report all 13 individual codes.
% 4 Participants did not answer this question.
When asked which types of studies should be replicated, 8 of the 25 participants responded that all studies should be replicated, and not just a specific subset (\textit{``I think it would be good to replicate all studies to see if their findings hold [...]''} -P12). %Some participants also had concerns or even completely disliked the idea about conducting replication studies. 
Regarding the question whether there should be more replication studies only one participant disagreed and stated that it is not worth replicating, while then explaining that the papers with the greatest impact should be replicated. Thus, all agreed that it is worth replicating. 
The person initially arguing against replication studies also mentioned the huge variance in user studies (noise) and claimed that therefore replication studies are more likely to find other results instead of confirming previous results (\textit{``Most aren't worth the time. The ‘noise' in most user studies is SO large that most folks know there will be differences.''} -P1). 

\section{Discussion}

Before we start discussing our results, we want to compare our work with a paper which was published while we revised our version based on reviewer feedback.

\subsection{Comparison with Schmüser et al. (2026)}
\label{PositionPaper}

Schmüser et al. (2026) \cite{schmuser2026position} also analyzed  Calls for Papers and identified replication studies in usable security and privacy. To do so, they examined the same conference venues as we did.
\ \\

\textbf{CfP / Our RQ1.} Schmüser et al. categorized venues based on the presence of explicit call for replication-related research in their Call for Papers. They cover the period between 2018 and 2026. Our analysis of the CfP pages also includes the years 2016 and 2017 which seem important as several venues started including the topic of replication in 2016 and in 2017: We see SOUPS began placing a strong focus on replications with their introduction of categories of replications (see Section \ref{CfP}), in 2017. A year later, SOUPS expanded the description of replications but removed the categories -- a relevant inside which is only discussed in our paper but is missing in \cite{schmuser2026position}. By starting our examination of the CfPs earlier, we also show that EuroUSEC (and USEC) explicitly encouraged the submission of replications as early as 2016 \footnote{USEC CfP 2016: \url{https://www.ndss-symposium.org/ndss2016/submissions/usec-workshop-call-papers/} Last accessed: August 1st 2026}$^,$\footnote{EuroUSEC CfP 2016: \url{https://www.ndss-symposium.org/ndss2016/submissions/usec-workshop-call-papers/} Last accessed: August 1st 2026}. In addition, USEC's CfP is not part of Schmüser et al.'s analysis. 

Furthermore, there are some issues with the statements of  Schmüser et al.: The authors classify CHI's CfP as ``[Replication] mentioned as acceptable''. However, this applies only to the subcommittee ``Computational Interaction''. which is not the subcommittee for usable security and privacy research which is the focus on their and our investigation.   For ICSE, both analyses find mentions of replications in the CfP for the years 2020 and 2021. However, Schmüser et al. also describe mentions in subsequent years through 2024 and refer to the reviewer guidelines. It is unclear, whether additional sources like the reviewer guide are included in the analysis of other venues' CfP. In our analysis, we strictly limit the analyses to the information provided in the CfP as it is presented to potential authors.
% interessant ist ob andere kategorien und ob grundsätzlich gleich und obs worth ist vor 2018 zu schauen, z.b. weil da mal mehr im cfp bei soups stand.
We also find encouragement of replications in an archived version of the CfP of EuroUSEC 2018\footnote{Archived EuroUSEC CfP 2018: \url{https://web.archive.org/web/20180625024845/https:/eusec.cs.umd.edu/#cfp}} while Schmüser et al. showing the venue's CfP starting encouragement in 2021.
Their remaining findings regarding the CfPs from the various venues are consistent with those from our analysis. \ \\

\textbf{Literature Search / Our RQ2.} Schmüser et al. \cite{schmuser2026position} used a keyword search in the title and abstract to look for the term ``replication'' over the past 20 years.    
They identified 14 replication papers\footnote{Note, Schmüser et al. also has three papers which are not in our list: Two papers published before 2016 (i.e., \cite{atwater2015leading,sotirakopoulos2011challenges}) and one paper which we excluded as  it did not constitute a user study (\cite{bird2020replication}).}, 12 of which appeared at SOUPS and two at EuroUSEC. No replications were found at the other venues while we found 11 at the other venues (e.g. \cite{busse2019replicationSnooze,naiakshina2020conducting,noah2025replication}). We can only speculate what is the reason for this discrepancy:  Based on the description of the search and the actual papers identified, it is unclear if the same venues as for the CfP were considered\footnote{We considered for RQ1 and RQ2 the same venues (SOUPS, EuroUSEC, USENIX Security, CHI, ICSE, NDSS, CCS, S\&P CSCW, EuroS\&P, WWW and PETS).}  %considered for the CfP analysis.} %have been searched (SOUPS, EuroUSEC, USENIX Security, CHI, ICSE, NDSS, CCS, S\&P CSCW, EuroS\&P, WWW and PETS) 
or if the search was limited to SOUPS and EuroUSEC.
Furthermore, it looks like that their search was limited to the title starting from 2018, since 2018, the SOUPS CfP suggests the use of the prefix ``Replication:'' in the title for replications submitted to the conference. This would explain why we identified two more  papers at SOUPS in the past ten years, which we consider replications but do not use this prefix in their titles \cite{al2018effectiveness,hansch2018programming}. %olor{red}{todo}. 
%It appears that the search was limited to the title for SOUPS starting from 2018. Since 2018, the SOUPS CfP suggests the use of the prefix ``Replication:'' in the title for replications submitted to the conference. 
The lack of 13 papers is a clear limitation of their work. % as we identified two papers published at SOUPS

%all identified papers from Schmüser et al. (2026) are included in our data set. 
Out of the 13 additional replication papers in our dataset, two contained the word ``replication'' in their title \cite{busse2019replicationSnooze,noah2025replication}, one paper contained ``Repli'' in its title \cite{mathis2021replicueauth}. Eight papers mentioned ``replication'' in their abstract (\cite{farzand2025scales,barber2025beyond,ray2021older,warberg2019can,ismail2017permit,naiakshina2020conducting,al2018effectiveness,hansch2018programming}). There is one paper which did not use any of the search terms in title or abstract (\cite{haring2023less}) but was found due to the incorrect search results from the ACM Digital Library (see Appendix \ref{App:LitSearch}). One more replication paper came to light through its mention in another replication paper \cite{naiakshina_if_2019}.
Thus, limiting the search to the term ``replication'' comes with an obvious limitation.
\ \\\\

\textbf{Further Results.} Note, Schmüser et al.  did not try to systematically categorize the papers they identified (our RQ3) nor tried to get insights from the authors (our RQ4). They only checked for each paper whether the findings could be confirmed or not. They found that many could not be confirmed. From the papers they identified, they conclude ``[..] that there is currently no consensus in the HCS community for how to conduct, describe,
and label replication studies.'' \cite{schmuser2026position}.

\subsection{Call for Papers - RQ1}

SOUPS is the first and only venue which provides additional information regarding replication papers. It defines replication as research that confirms, questions, or clarifies the findings of prior studies. Notably, the call does not specify whether the research problem must be defined identically to that of the original study. Thus, other than in the framework, the focus is on the findings, rather than on the studied problem. %This definition encompasses a broad spectrum of work, including studies that vary in domain, methodology, and analysis, categorized as Group 8 within the framework. 
%While the SOUPS call for papers emphasizes that replications of important and influential research are particularly valuable, it offers no guidance on how such importance or influence should be evaluated. Furthermore, an additional rationale for conducting replications, namely, the presence of surprising or unexpected results in the original study, is not addressed, despite its potential value to the research community. 
Furthermore, there is no guidance about information which needs to be provided for replication papers as it only requires to have the term in the title.

Moreover, all other CfPs which mention replication completely omit any description of what they deem valuable contributions for replication studies. Interestingly, the subcommittee for privacy and security at CHI does not mention replication papers at all. 

\subsection{Lessons Learned from Classification - RQ2 and RQ3}
In this work, we have made a first attempt to understand and classify the current state of replications in the field of usable security and privacy. For this purpose, we adapted a replication framework by~\cite{olszewski_sok_nodate} to enable us to classify replications of user studies. We summarize our lessons learned classifying the 24 papers regarding replications in the usable security and privacy context. % we identified in a systematic literature search. 

\ \\\textbf{General Lessons Learned.} %It is likely that the authors of these 24 papers had a less systematic understanding of a replication paper as compared to the systematic approach of the replication framework. The responses to our replication author survey confirm this. 
%Furthermore, as seen in the survey answers and also f
%From reading and classifying the papers, 
Due to the lack of %common understanding of a replication and 
clear requirements for replication papers, it was sometimes challenging to categorize papers according to the framework. However, we believe that the framework is an important step forward towards a common understanding of replication papers.  Several papers were not mainly focusing on validating the results from previous papers but instead focused on extending the original paper by answering related or follow-up research questions. This is in line with the findings from Hornbæk et al.~\cite{hornbaek_is_2014} from 2014. %: a majority of authors of replication studies had not the replication of the original study as their main goal, but to research additional topics.

\ \\\textbf{Lessons Learned Regarding the ``Root Problem''.}
According to the framework, a precondition for being a replication is to study the same problem.  However, the 24 papers we investigated %(as well as several we excluded in the process before) 
were not very clear about this. Statements made were, ``we replicated the paper'' or ``we replicated the methodology''. Therefore, we had to %be more open and could not require authors to explicitly state that the problem / research questions are the same, but rather 
check ourselves whether the addressed problems were indeed the same across the original and replication studies, even if it was not stated specifically. %\textcolor{red}{room for interpreation in limitations aufnehmen}
Furthermore, we noticed that depending on the phrasing of the problem in the original paper (being broader or very specific), related problems can be considered as replication or not. This is a limitation of the original framework and our adaption, that results in challenges when classifying past papers. However, for future papers this could be addressed by allowing authors of replications to explain why they believe their research counts as replication.

%* Challenging if original study has more than one domain / method / analysis.
\ \\\textbf{Lessons Learned Regarding Mixed-Methods.} 
One challenge that we encountered repeatedly, is that many replication efforts are not confined to a single methodological approach. Instead, they rather collect quantitative and qualitative data in one study, implying the need for different analysis approaches in the same study. In several cases the original and the replication paper run different types of user studies. Thus, the framework should be more specific regarding whether it classifies entire papers or each method/analysis is treated separately. We are in favor of the second approach. %comprise multiple studies that employ distinct methods. %Consequently, the comparison with the original investigation is not always straightforward, especially when one of the included studies adopts a different methodology while another does not.

\ \\\textbf{Lessons Learned Regarding Several Replications for One Paper.} Related to the previous lessons learned, we noticed that the framework should be more specific regarding replications which are not the first of the original paper. We propose to provide classification of replications in relation to the original paper and if applicable for each of the existing replications separately. 
 original analysis what to consider as same/different. (e.g. same code book or create code book like the original)

\ \\\textbf{Lessons Learned Regarding ``Same'' versus ``Different''.}
Another issue concerns the precise interpretation of the phrase ``same analysis.''  For example, when qualitative data are coded with a codebook, it is unclear whether a replication requires constructing an identical codebook, mirroring every step of the original development, or merely re‑applying the original codebook to a new data set.  Using the pre‑existing codebook facilitates direct comparison of results.  However, this also propagates any latent deficiencies or biases embedded in the original codebook, underscoring the importance of conducting a replication that explicitly evaluates and, if necessary, corrects such issues. In quantitative research for example, after a structural equation model (SEM) has been constructed in an original study, should the ``same analysis'' mean that a replication follows the same approach in constructing a SEM with new data or should it use newly collected data to validate the existing model through a model-fit evaluation. 
%\textcolor{red}{Benjamin kannst du noch was aufnehmen zur quantiativen analyse model selbst erstellen wie original oder validieren}. 
%Here, clarification is needed based on a discussion within the usable security and privacy community. 
We recommend authors of replication papers to state whether they consider it to be different and justify their decision for different / same.

We also identified small differences (e.g. just changing from a 5-point to a 7-point scale) as well as big differences in both domain and methodology. Thus, it can be discussed to distinguish same and differences by at least mentioning the amount of differences.

\ \\\textbf{Lessons Learned Regarding Partial Replications and Replication Extensions.} We saw both types several times. We propose to focus on (the parts of) the problem which are the same, when categorizing replications. Extensions should be treated as new (follow-up) research instead.

\subsection{Motivation to Replicate - RQ 4}

The aim of the survey was to understand the motivation of why researchers replicate studies. We found certain themes to be the main driver for replications in usable security and privacy: to re-evaluate study findings at a later point in time or to validate findings for a different sample (e.g. different country). This is in line with our findings from the classification of replication studies.

%Deren Motivation ist validierung über Zeit, andere Länder -> was passt zum Rest
%Ist in line -> temporal und contextual changes was sie sagen und was wir gefunden haben

%% hier hin verschoben
\subsection{Limitations}
One limitation of the systematic literature search is the focus on the 13 venues and the search terms ``replication'', ``reproduce'' and ``repeat''. For venues whose subject area is not exclusively security and privacy, we also used the two terms ``security'' and ``privacy'' in the search. The main limitation of the categorization is that we took a worst case approach as already small changes and a change of just one aspect of the study were assigned to ``different''. %Furthermore, the papers were not always clear about the actual differences.
%The main limitation of the search for replication papers of most cited papers is the choice of search terms. There might be other terms like re-validate and follow-up which would have resulted in a few more papers. 
The main limitation of the survey with authors is that only some of the authors answered our survey. %A wider sample might have led to more positive or negative impressions not captured in our study. Nevertheless, the surveyed authors have already noted critical aspects that can provide meaningful improvements and  thoughts for the community. 
%In addition, the authors participated in an online survey and not in a detailed interview. Therefore, the results can only be considered as a small insight into the motivation for  a replication and should be examined even more thoroughly in the future in a separate study.
% \textbf{\textcolor{red}{xxx replicated the paper xxxx }}published at IEEE S\&P. Note, this paper is not in the list of replications identified in Section \ref{SearchConf} because the authors of this paper used in their title and abstract only the term re-validate \textcolor{red}{ggf als limitation in Section \ref{SearchConf} [oder sammeln wir die limitations im discussion kapitel am ende] aufnehmen, dass wir eben nur nach den drei papern gesucht haben.} \textcolor{red}{noch was sagen, was da gemacht wird?}

% \textcolor{red}{das thema noch aufnehmen:?}Therefore, we had to be more open and could not require
% authors to explicitly state that the problem / research ques-
% tions are the same, but rather checked ourselves whether this
% was the case although it was not stated specifically. room for
% interpreation in limitations aufnehmen

 %Dadurch, dass es keinen einheitlichen Begriff gibt, hätten weitere Begriffe wie "follow-up" etc. möglicherweise weitere Paper gezeigt.

\section{Proposal for Concrete Guidance for Replication Papers}

%\textcolor{red}{TODO: hier fehlt irgendwie was, dass relativ chaos, dass es helfen würde klarer zu sagen was für art von replications gesucht werden im CfP und den autoren klarer zu sagen, was zu tun ist }
%Our findings indicate that replications are still relatively rare, even for high-profile papers that are among the top-cited papers of usable security and privacy venues. Therefore,
We used all our findings, to propose clear guidance for replication papers in usable security and privacy. Our proposal is based on the
SOUPS CfP  (see Figure \ref{fig:CFP_proposal}) and includes a template (see Table \ref{tab:cfptemplate}) in latex to indicate differences between the original work and the replication. The table is structured according to the framework from this paper. Note, the table indirectly indicates whether it is only slightly different or whether there a big differences in domain and method as there would be a longer list of aspects being different.
\ \\\\
The guidance  addresses the following aspects:
\begin{itemize}
    \item Authors   state whether it is a replication, a partial replication, or an replication extension. 
    
    \item Authors  justify the replication as such.

    \item Authors  state what is the problem addressed by the original paper and what is the problem addressed by their replication (which might not constitute the entire contribution of a paper in case of a Replication-Extension paper).
    \item Authors explain the involvement of the authors of the original paper (e.g. getting access to additional information / data) and in case at least one author of the original paper is an author of the replication paper, this should be discussed. 
    \item Authors explain and justify their adaptions to the original domain, method, and analysis\footnote{Reviewers are asked to verify these statements.}.
    \item Authors compare their results with the results of the original paper. 
    \item Replication papers should be self-contained and comprehensible without requiring prior knowledge of the original work.
    
\end{itemize}
Figure \ref{fig:CFP_proposal} contains a proposal how to address this guidance in a Call for Paper. 
%First of all, for future replication papers, 
%Furthermore, we propose the  framework from this paper as clear guidance for authors to classify and justify their replications themselves and reviewers then only need to verify whether they agree or not. 

\begin{figure*}[!ht]
  \centering
  \includegraphics[width=\textwidth]{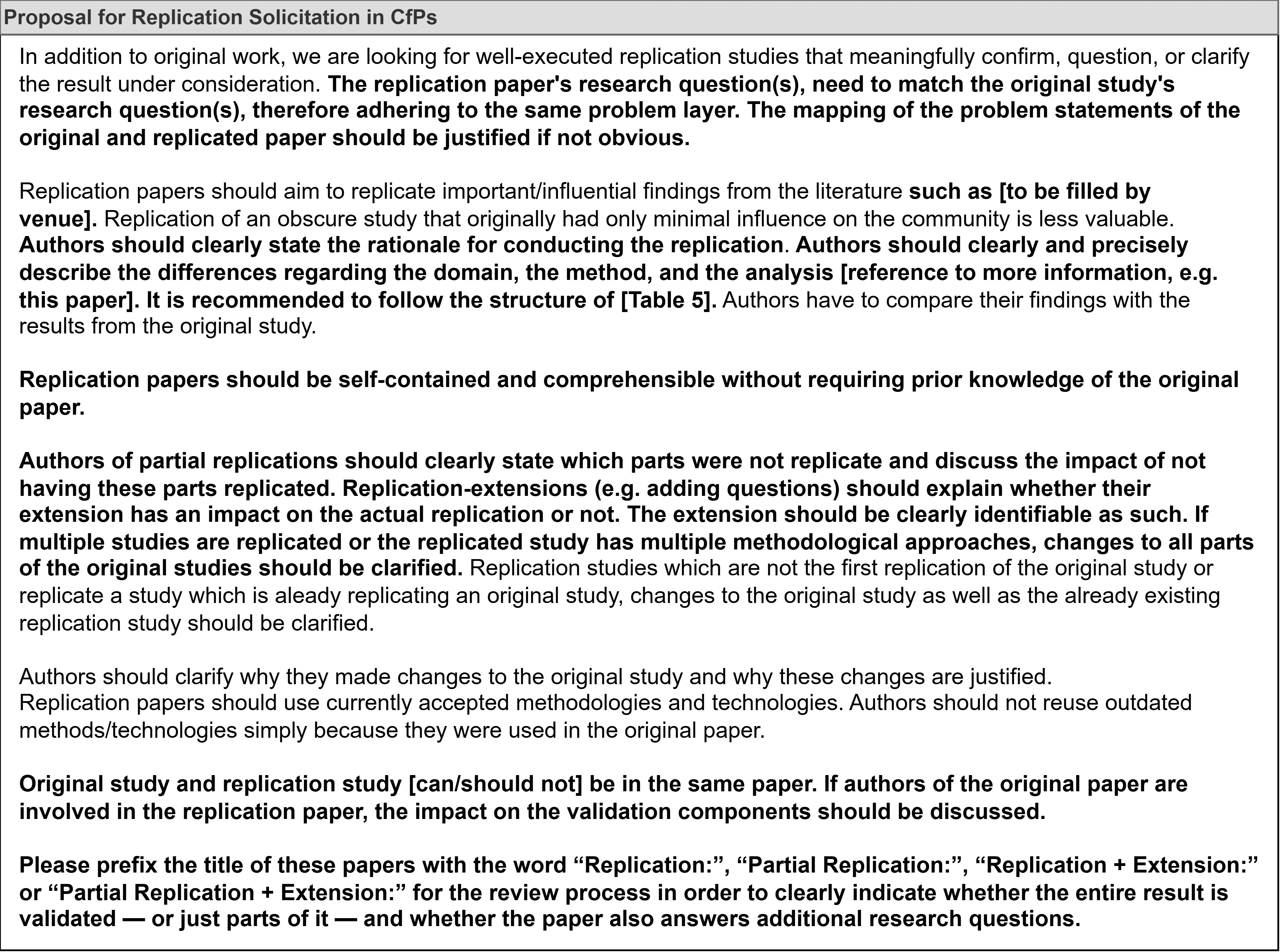} 
  \caption{Proposal for Replication Solicitation in Call for Papers.}
  \label{fig:CFP_proposal}
\end{figure*}

Chairs, authors and reviewers in the usable security and privacy context%\footnote{Note, this also holds for security and privacy in general. }%Therefore, we would like to encourage other security and privacy subdisciplines to develop similar proposals as we assume to see similar phenomena when replicating other types of security and privacy papers. }
-- as well as the entire community -- benefit from such  guidance.
To further support them, we propose a latex template to specify the differences regarding domain, method, and analyses in a table in replication papers (see Table \ref{tab:cfptemplate}\footnote{For the latex template go to \url{https://files.secuso.org/secuso/2026-08-Replication_Practices_Usable_Security_Privacy/} Last accessed: August 1st 2026}). %Our proposal would also support authors and reviewers. 

 % A replication study should be self-contained and comprehensible without requiring prior knowledge of the original work. Therefore, including a comprehensive description of the original study in the methodology section appears advisable.
%\textcolor{red}{intervention - replication}
Among the replication papers included in our review, certain conventions emerge in some instances that we consider worth following.
The methodology section can be divided into three sections: (1) the original study design, (2) deviations from the original protocol, and (3) extensions to the original study (if applicable).
Modifications to the original study should be explicitly justified, and their potential impact on the results should be assessed. This applies not only to questionnaires and study protocols but also to recruitment practices.
In addition, it is useful to present the original results in the ``Results'' section (for example, in the form of a figure or table) even if no additional statistical analyses are performed using the original data.

Overall, we enable best practices on how to conduct, describe,
and label replication studies and thereby address the lack of consensus in the community as identified by  \cite{schmuser2026position}. Schmüser et al. (2026) also recommend to develop corresponding guidance.  %\textcolor{red}{hier schreiben dass das auch zur motivation hilft welche durchzuführen, siehe survey} \textcolor{blue}{In addition, this comparison table provides authors of potential replication studies with a clearer understanding of the key differences. This structured overview supports informed decisions about whether a replication warrants execution in its current design.}

% research community. 

%We recommend venues that publish usable security and privacy research  to provide clear guidance for replication papers in this context. To aid venue organizers, we have developed a proposal based on the SOUPS CfP and our own findings (see Figure \ref{fig:CFP_proposal}). %which can be found in Appendix \ref{app:cfpproposal}. 
%Those who wish to develop their own text based on our findings should -- in particular -- take Figure \ref{validityTree2} and Table \ref{tab:cfptemplate} into consideration. % \textcolor{red}{ausbauen und sagen, dass das authors und reviewers hilft; , e.g. to check whether all information is available to understand the differences.}
%\textcolor{blue}{In addition, there could be a clearer definition of what venues consider to be valuable contributions through replication studies. This additional clarity could motivate more researchers to conduct and submit replication studies, knowing that the respective venue values the type of replication study in question.}

\begin{table*}[!ht]
\centering
\footnotesize
\renewcommand{\arraystretch}{0.85}
\setlength{\extrarowheight}{-0.5pt}
\setlength{\aboverulesep}{0.2pt}
\caption{Comparison of original studies and the replications conducted in this work. Examples are provided as guidance and are not intended to be coherent across the different types of information that should be provided. If it is necessary to go into more detail regarding differences in specific parts of the study (e.g. the questionnaire) it may prove useful to present these differences separately in a second table.}
\label{tab:cfptemplate}
%\begin{longtable}{cllp{6.5cm}}
\begin{tabular}{cp{1cm}p{2cm}p{11.5cm}}

\toprule
\multirow{2}{*}{\shortstack{\textbf{Scope of}\\\textbf{information}}} & \multicolumn{2}{c}{\textbf{Information to Provide}} & \multirow{2}{*}{\textbf{Explanation}} \\
 & \textbf{\textit{Framework Layer}} & \textbf{\textit{Type of Information}} & \\
\toprule
%\endfirsthead

% \\
%\toprule
%\endhead 

%\multicolumn{4}{r}{Continued on next page}\\
%\endfoot

%\bottomrule
% \\
%\endlastfoot

\multirow{42}{*}{\rotatebox{90}{\shortstack{\textbf{Each Study}\\\textit{(Original, Replication, \& all potentially relevant previous papers)}}}} & Problem & Research Questions & The research questions provided to frame the exact problem the study is about. Either repeat in verbatim from the text or refer to exact location in your manuscript. Typically, from these it should be clear why the original and replication studies are investigating the same problem.\\
\cmidrule{2-4}
 & Domain & Sample Size & The number of participants for all study conditions. Example: \textit{$n=10$ Treatment (one excluded during analysis), $n=11$ Control} \\
\cmidrule{3-4}
 &        & Type of participants & A description of the individuals that made up the sample. Examples: \textit{Computer science university students}, \textit{Elderly residing in a care home} \\
 \cmidrule{3-4}
 & & Recruitment & The methods of participant recruitment. Examples: \textit{Online panel (Prolific)}, \textit{Snowball-sampling through network of researchers}\\
 \cmidrule{3-4}
 &        & Compensation & The type and value of compensation the participants received. Examples: \textit{2€ through panel provider}, \textit{10€ voucher for ice cream shop across on the university campus} \\
 \cmidrule{3-4}
  &        & Used Devices \& Services & The domain used to investigate the research questions is best framed based on the devices and services used as part of the investigation. Examples: Service -- \textit{Google Activity Dashboard (as of 2026-02-19)}, Hardware -- \textit{Hololens 2 AR HMD}, Self-created Research Artifact -- \textit{Own web extension browser add-on used in Firefox ($<$link to repository with add-on$>$)} \\
\cmidrule{3-4}
 &        & Versions of Used Software & Different versions of software can change feature sets and appearances dramatically. The goal should be to be as precise as possible. Example: \textit{Firefox browser (version 147.0.4, used in Private Window, start page change in setting to tutorial for research artifact, no other add-ons installed except own research artifact add-on); macOS (version 15.7.4, fresh install with only configuration change being fixed dark mode)}\\
\cmidrule{2-4}
 & Method & Study Setting & The circumstances in which the participants worked on the  study. Examples: \textit{Lab Study (laptop provided by researchers with aforementioned software)}, \textit{Field Study (participants' own laptop with the aforementioned software being installed on their own)} \\
\cmidrule{3-4}
 &        & Study Conditions & The different conditions in the study. Example: \textit{Treatment (browser add-on), Control (same browser without add-on)}\\
\cmidrule{3-4}
 &        & Measured Data & A detailed enumeration of all measured data and means of collection. Example: \textit{Qualitative think-aloud recordings with subsequent transcription, quantitative data (SA-6, ATI) through survey}\\
\cmidrule{3-4}
 &        & Other noteworthy aspects & List any other noteworthy aspect about the used methodology, such as the use of deception. Example: \textit{Study purpose was not disclosed beforehand but only in a debriefing at  the end of the study}.\\
\cmidrule{2-4}
 & Analysis & Analysis Methods &  List all the analyses you performed. For quantitative analyses list all IVs and DVs. Example: \textit{Regression analysis (DV: Trust, IVs: study condition, SA-6, ATI); Qualitative thematic coding for think-aloud data}\\
\midrule
\multirow{5}{*}{{\shortstack{\textbf{Once}\\\textbf{for Paper} \vspace{0.5cm}}}} & \multicolumn{2}{p{3cm}}{Key differences between original and replicated studies} & This should provide a brief overview of the most important differences between the original and the replication studies. Additional information that should be added is how much time has passed between the studies and potentially which other previous papers are relevant.% (e.g., in case the replicated paper is itself a replication of an earlier paper). 
\\
\bottomrule

\end{tabular}
%\end{longtable}
\end{table*}

 \section{Conclusion and Future Work}
We investigated the  nature of replications published in the domain of usable security and privacy. %. We considered the mentioning of replications in CfPs from 13 venues. 
We searched for replications in 13 venues from 2016 to 2025 and identified 24 replications of which 11 were published at SOUPS, which regularly mentions the acceptance of replications in their CfPs. We subsequently applied an adapted replication framework originally proposed by Olszewski et al.~\cite{olszewski_sok_nodate} to categorize these 24 papers. %Furthermore, we investigated whether the most cited SOUPS papers were replicated: 3 out of the 19 were replicated. 
%Our analysis indicates that there are only very few replications at all and t
Those replications we found are no exact replications. 
Most replications are substantially different wrt. the domain, the method, and the analysis. % modified from the original designs.
To get more insights in the actual work on replicating research, we surveyed the authors of the identified replication papers. % to complete an online survey, in which we asked about their motivation and general opinions on replications. We received 25 responses in total. Frequently cited reasons for performing a replication were validation, observation of changes over time and increased efficiency afforded by reusing existing materials and methodologies. While the majority of respondents agreed that every type of study should be replicated, a minority expressed concerns that the high variability inherent in user‑study data could undermine the feasibility of meaningful replication in this domain.

%In a final step of the study, we investigated whether the most cited SOUPS contributions from the past decade had been subjected to replication. We identified 19 SOUPS papers cited over 150 times, extracted all their citing articles via Google Scholar, and after reviewing 2,108 candidates that mentioned related keywords in their text, found only three replications.
From the various investigations, we deduced lessons learned for applying the adopted framework as well as recommendations on how to specify in Call for Papers replication studies to support authors, reviewers, and the community as such. % incetivize replications including a first proposal for concrete guidance.
%
%While examining the replications, we found that most studies did not provide a concise summary of the modifications relative to the original work; therefore, we recommend including a standard table in future papers that explicitly lists the differences (for an example see Table~\ref{tab:freq} in the Appendix).%\textcolor{red}{Tabelle von Peter verlinken oder referenzieren}
%

Future work is mainly to apply the framework and the recommendation for replications and improve the proposal based on these experiences. Regarding the actual paper writing, we want to start a discussion with the community on what to expect in the related work section of replication papers. % conducted post having submitted this paper. %, and finding ways -- such as poster presentations or workshop papers -- to start a discussion with the community to answer the collected questions. 
%Future work should focus on refining the applied framework through further application, iterative adaptation, and explicit resolution of remaining ambiguities. Expanding the literature search to incorporate broader terminologies, such as “validate” and “follow‑up”, would likely capture a wider range of replication efforts.
 
%- was gemacht. 

%Conclusion: es muss was getan werden. Diverse Recommendations abgeleitet.

% - Future Work: collecting experiences with applying our approach when applying

% - man könnte sagen, man teilt das nochmal auf, dass man jedem paper mehrere typen zuordnen kann, weil man es eben aufteilt und auch wenn es mehr originalpaper gibt.
 
 %Researcher fragen, woran es liegen könnte? z.B. die fragen die mehr als 5 paper hatten bei einem der venues aber noch keine replication. 

% \textcolor{red}{in future work aufnehmen, z.b. bei diesem neuen workshop zu diskutieren oder Poster Sessions bei den 13 venues, die ganzen punkte, die oben zur diskussion stehen}


\section*{Acknowledgments}
This work was funded by the Topic Engineering Secure Systems, subtopic 46.23.01 Methods for Engineering Secure Systems, of the Helmholtz Association (HGF) and supported by KASTEL Security Research Labs, Karlsruhe.

\newpage


\bibliographystyle{IEEEtran}
\bibliography{Replication}

\appendix
\renewcommand{\thesection}{\arabic{section}}
\renewcommand{\thesubsection}{\arabic{section}}
\subsection{Literature Search Details}
\label{App:LitSearch}
For all searches, the search period was limited to 2016-2025.

\subsubsection{ACM Digital Library: SOUPS, CHI, CCS, ICSE, CSCW, WWW}
\label{app:acm}
Use of the Advanced Search with search queries \url{https://dl.acm.org/search/advanced?expand=dl&text1=0&editQuery=true} (last accessed: 18.02.2026). To use the queries below, extend the option ``View Query Syntax'' and paste them into the ``Edit Query'' text box. Set the custom publication date range to \textit{Jan 2016 to Dec 2025}.

To limit the search to a specific venue, use the sidebar menu (\textit{Publications / Proceedings Series}) on the results page. Search queries used for the ACM Digital Library search:

\begin{itemize}
    \item Title:(Replication AND (Privacy OR Security)) OR Abstract:(Replication AND (Privacy OR Security)) OR (Title:(Replication) AND Abstract:(Privacy OR Security)) OR (Abstract:(Replication) AND Title:(Privacy OR Security))
    \item (Title:(Reproduce OR Repeat) OR Abstract:(Reproduce OR Repeat)) AND (Title:(Security OR Privacy) OR Abstract:(Security OR Privacy)) AND Title:(!(Replication)) AND Abstract:(!(Replication))
\end{itemize}

\textbf{CHI, ICSE, CSCW, WWW}: Use of full search query.

\textbf{CCS}: Due to the topic of the venue, the terms "Privacy" and "Security" were excluded int his search.

\textbf{SOUPS}: In order to access search for SOUPS publications visit: \url{https://dl.acm.org/search/advanced?AllField=0&ConceptID=119574&expand=all&target=advanced&editQuery=true&fillQuickSearch=false} The proceedings and abstracts of the SOUPS papers between 2016 and 2025 were hosted on the ACM DL. Since the papers themselves were not hosted those entries did not include results due to the inclusion of AI summaries. The search query did also not include the keyword "Security" and "Privacy".

\textbf{ACM Digital Library Search Issue}: During our literature search at the beginning of 2026 we ran into an issue with the ACM Digital Library search function. When searching by title and abstract, the search engine also included the AI-generated summary in its search scope. This resulted in papers being included in the search results which had the search term neither in the title nor abstract but in the AI-generated summary of the paper. We contacted ACM who acknowledged the issue but could not easily fix it.

\subsubsection{IEEE Xplore: S\&P and EuroS\&P}
\label{app:ieee}
Use of Advanced Search through option \textit{Command Search} \url{https://ieeexplore.ieee.org/search/advanced/command} (last accessed 18.02.2026)
Title and Abstract were searched for the terms Replication, Reproduce and Repeat. 
The following search queries were used for EuroS\&P:
\begin{itemize}
    \item ("Publication Title":""European Symposium on Security and Privacy" AND ("Abstract":Replication OR "Document Title":Replication))
    \item ("Publication Title":""European Symposium on Security and Privacy" AND (("Abstract":Reproduce OR "Document Title":Reproduce) OR ("Abstract":Repeat OR "Document Title":Repeat)) AND NOT("Abstract":Replication OR "Document Title":Replication))
\end{itemize}

The resulting list included papers from workshops which were part of the conference.
For S\&P the same search queries were used except for the change in publication. 
The list for S\&P did not include any workshops (except EuroS\&P workshops).

\subsubsection{USEC, EuroUSEC, AsiaUSEC}
\textbf{USEC}: 
Due to changes in the website structure for USEC over the years, the approach differs from year to year. In all cases did the keywords include: "repli", "repro" and "repea".
\begin{itemize}
    \item 2016-2019: manual keyword search in each paper's title and abstract
    \item 2020: USEC did not take place.
    \item 2021-2023: Keyword search through USEC symposium website  source code (hosted on NDSS and includes title and abstract)
    \item 2024, 2025: Keyword search after scraping title and abstract from NDSS website
\end{itemize}

\textbf{EuroUSEC}: EuroUSEC changed publisher multiple times over the years. Based on the publisher different search approaches were chosen. Titles and abstracts were searched for the keywords: "repli", "repro" and "repea".
\begin{itemize}
    \item 2016-2018: NDSS, manual search of each paper's title and abstract.
    \item 2019. 2020: Title of the published papers taken from the proceedings and searched for on Google Scholar. Paper manually searched for keywords in title and abstract. 
    \item 2021-2024: ACM, search query according to see Appendix \ref{app:acm}. "Privacy" and "Security" were excluded.
    \item 2025: IEEE, search query according to Appendix \ref{app:ieee}.
\end{itemize}

\textbf{AsiaUSEC}: Did take place only once in 2020. It was a co-event at Financial Cryptography and Data Security. The titles of all papers published at AsiaUSEC were taken from the Financial Cryptography and Data Security proceedings and searched on Google Scholar. A manual keyword search was performed for each paper only considering title and abstract. The used keywords were: "repli", "repro" and "repea".

\subsubsection{Use of own scraping scripts: PETS, NDSS, USENIX Security}
The three venues PETS, NDSS as well as USENIX Security were not searchable through a publisher search engine. Scraping scripts\footnote{For the scripts for NDSS, PETS and USENIX Security see \url{https://files.secuso.org/secuso/2026-08-Replication_Practices_Usable_Security_Privacy/} Last accessed: August 1st 2026} were written to download title and abstracts for all papers published at those venues between 2016 and 2025. In some cases single years did not post abstracts on their website. In those cases the PDF files had to be searched manually, focusing on title and abstract.

All titles and abstract were keyword searched with the expressions: "repli", "repro" and "repea" in order to locate all mentions of all variations of the word replicate, reproduce and repeat.

\subsubsection{Search with Google Scholar}\label{mostcitedsearch}
The following steps were taken for each of the 19 papers:
\begin{enumerate}
    \item Search for the title on Google Scholar
    \item Click on ‘cited by'
    \item Select option ‘Search within citing articles'
    \item Enter search term: "replication" OR "replicate" OR "replicating" OR "reproduce" OR "reproduction" OR "reproducing” OR"repeat" OR "repeating" OR "repetition"
\end{enumerate}
Google scholar's search function includes the entire text of the papers, and, to our knowledge, there is no possibility to restrict the search to abstract and title, which explains the large amount of papers.

\subsection{Author Survey Questions}
\label{appendix:authorsurvey}

\begin{itemize}
    \item[1.] Why did you decide to replicate a user study?
    
    \textit{[open ended response]}
    
    \item[2.] Why did you decide to replicate your specific user study? If you replicated more than one study, than choose your most recent one.
    
    \textit{[open ended response]}
    
    \item[3.] Do you think there should be more replication studies?
    
    \textit{[Yes/No]}
    
    \item[4.a] [if ``Yes'' in 3.)] Which type of security/ privacy related user studies should be replicated (more often)?
    
    \textit{[open ended response]}
    
    \item[4.b] [if ``No'' in 3.] Why do you think it is not necessary to have more replication user studies in usable security and privacy?
    
    \textit{[open ended response]}
    
\end{itemize}

\onecolumn

\end{document}